\documentclass[fleqn,usenatbib]{mnras}

\usepackage{newtxtext,newtxmath}

\usepackage[T1]{fontenc}
\usepackage{ae,aecompl}
\usepackage[usenames, dvipsnames]{color}

\usepackage{graphicx}	
\usepackage{amsmath}	
\usepackage{amssymb}	
\newcommand{\angstrom}{\mbox{\normalfont\AA}}
\usepackage{diagbox}
\usepackage{booktabs,multirow,array}
\usepackage[table]{xcolor} 

\title[Covariance-regularized reconstruction of data cubes]
{Covariance-regularized reconstruction of data cubes in 
integral field spectroscopy and application to MaNGA data}

\author[D. Liu et al.]{
Dou Liu$^{1}$, Michael R. Blanton$^{1}$, David R. Law$^{2}$\\
\\
$^{1}$Center for Cosmology and Particle Physics, Department of Physics,
New York University, 4 Washington Place, New York, NY 10003, USA\\
$^{2}$Space Telescope Science Institute, 3700 San Martin Drive,
Baltimore, MD 21218, USA\\
}

\date{Accepted XXX. Received YYY; in original form ZZZ}

\pubyear{2015}

\begin{document}
\label{firstpage}
\pagerange{\pageref{firstpage}--\pageref{lastpage}}
\maketitle

\begin{abstract}
Integral field spectroscopy can map astronomical objects spatially and 
spectroscopically. Due to instrumental and atmospheric effects, it is 
common for integral field instruments to yield a sampling of the sky 
image that is both irregular and wavelength-dependent. Most subsequent
analysis procedures require a regular, wavelength independent sampling 
(for example a fixed rectangular grid), and thus an initial step of 
fundamental importance is to resample the data onto a new grid. The best 
possible resampling would produce a well-sampled image, with a resolution 
equal to that imposed by the intrinsic spatial resolution 
of the instrument, telescope, and atmosphere, and with no 
statistical correlations between neighboring pixels. A standard 
method in the field to produce a regular  set of samples from an 
irregular set of samples is Shepard's method, 
but Shepard's method typically yields images with a degraded 
resolution and large statistical correlations
between pixels.  Here we introduce a new method,  which improves on 
Shepard's method in both these respects. We apply this method
to data from the  Mapping Nearby Galaxies at Apache Point Observatory 
survey, part of Sloan Digital Sky Survey IV, demonstrating a 
full-width half maximum close to that of the intrinsic 
resolution (and $\sim$ 16\% better than Shepard's method)
and low statistical  correlations between pixels. 
These results nearly achieve the ideal resampling. This method can 
have broader  applications to other integral field  data sets and 
to other astronomical  data sets (such as dithered images) with 
irregular sampling.
\end{abstract}




\section{Introduction}
Integral field spectroscopy (IFS) yields a rich set of information
about extended objects such as galaxies.  Modern facilities for 
performing IFS include the Multi Unit Spectroscopic Explorer \citep[MUSE;][] {Laurent2006}, the Keck Cosmic Web Imager \citep[KCWI; ][]{Morrissey2012}, and several efforts to create large samples of nearby galaxies,
such as the Calar Alto Legacy Integral Field Area 
Survey \citep[CALIFA;][]{Sanchez2012}, the Sydney-AAO Multi-object IFS \citep[SAMI;][]{Croom2012} and Mapping Nearby Galaxies at Apache Point Observatory  \citep[MaNGA;][]{Bundy2014}.

Here, we consider the MaNGA survey, one of three core programs in 
the fourth-generation Sloan Digital Sky Survey \citep[SDSS-IV;][]{Blanton2017}. MaNGA uses integral field units 
(IFUs) consisting of optical fiber bundles to obtain spectra 
across the face of a sample of 10,000 low redshift galaxies, 
making it possible to map the spectroscopic properties of
galaxies and to interpret this spectroscopy
in terms of two-dimensional maps of stellar age, gas phase
and stellar phase elemental abundances, star formation histories,
and kinematics.

The MaNGA fiber bundles are arranged in a hexagonal grid with a
separation of 151 microns. Each fiber has a cladding and buffer
in its outer annulus and has an active core size of 120
microns, corresponding to 2 arcsec in the Sloan Foundation Telescope
focal plane.  This
configuration results in an effective filling factor for the active
cores of 56\% \citep{Law2015}.  With a typical seeing full-width
half-maximum (FWHM) of 1.5 arcsec at Apache Point Observatory (APO), a
single observation undersamples the sky image considerably. Thus, to
increase the sampling of the point spread function (PSF) and to avoid
sampling irregularities, MaNGA uses dithered observations. Ideally,
each observation consists of three dithered exposures, performing a
set of subsequent exposures on each side of an equilateral
triangle. This pattern of observations leads to a finer, but still
hexagonal, pattern on the sky at the guiding wavelength. In the ideal
case when all observations are taken at the same hour angle, the
dithers produce a hexagonal pattern at all wavelengths, with
an overall wavelength-dependent shift of the hexagonal pattern on the
sky due to chromatic differential refraction by the
atmosphere. If the observations are taken at different hour
angles, the set of dithers will not produce a perfect hexagonal
pattern of samples at all wavelengths.
Thus, generally speaking, the MaNGA data produce an
irregular sampling of fluxes on the sky that varies with wavelength.

Most analysis techniques are designed to handle regular, usually 
rectangular, grids. Except for analyses involving a full 
``forward modeling'' of  the MaNGA data, resampling the fluxes 
onto such a regular grid is important. The resulting product is 
referred to as a data cube. These data cubes are the primary product
of the MaNGA data reduction pipeline (DRP; \citealt{Law2016}). 
The DRP extracts the
spectra and calibrates them, producing row-stacked spectra (RSS) --- 
one spectrum per fiber all interpolated onto the same wavelength grid.
At each wavelength, MaNGA then uses a modified version of Shepard's
method (\citealt{Shepard1968}) 
to resample the RSS data for that wavelength onto a rectangular 
grid.  This method, which is a flux-conserving variation of 
Shepard's interpolation method, is widely adopted because of its 
simplicity and robustness \citep{Yang2004}, for example in the 
CALIFA pipeline \citep[e.g.][]{Sanchez2012}.

In this paper we reconsider this choice of resampling method 
and propose an alternative. Before beginning, we ask: what 
is the best performance we should expect from a resampling algorithm? 
Here we define this ``best performance'' as being equivalent to 
actually sampling the real sky with the real instrument at the 
locations of every grid point in the desired resampling. This
process would produce an image with a resolution equivalent to 
the intrinsic spatial resolution of the instrument, telescope, 
and atmosphere, and zero statistical correlations between the
errors in each pixel. This definition means we are not attempting 
to deconvolve the intrinsic resolution --- just to resample 
with as little  loss of resolution or introduction of 
statistical correlations as possible.

Relative to this ideal, Shepard's method suffers two problems.  First,
it broadens the final reconstructed PSF substantially. In the cases we
consider in this paper, the FWHM of the final PSF can be 20\%
broader than the intrinsic resolution.
Second, the errors in pixels in
the reconstruction are highly correlated with each other. In the cases
we consider in this paper, for Shepard's method pixels separated by 1
arcsec can have correlation coefficients between their errors as high
as 0.70, complicating any correct analysis of the images that 
accounts for errors. These deficiencies
drive us to find a way to improve the performance of the resampling.

To do so, we make use of the techniques developed in a different context
by \citet{Bolton2010}, which they term ``spectroperfectionism.'' This name
alludes to the fact that they are performing a spectroscopic extraction.
However, we prefer the name ``covariance-regularized reconstruction'' as 
more generally applicable, since the approach of \citet{Bolton2010} is
appropriate for a number of other contexts beyond spectroscopic
extraction or (as applied here) integral field spectroscopy image
reconstruction.

We will make some important distinctions between different terms 
in this paper to avoid confusion: 
\begin{itemize}
    \item {\it Kernel}: The response $K(x,y)$ to a delta-function 
    source of 
    the atmosphere, telesecope, and instrument combined. In the 
    specific case of MaNGA, the kernel is the atmospheric seeing, 
    convolved with the telescope's optical response,  convolved 
    with the fiber profile (which we approximate as a 2 arcsec 
    diameter top hat).
    \item {\it Kernel-convolved image}: The actual or model image 
    on the sky convolved with the kernel.
    \item {\it Shepard's image}: The resampling resulting from 
    Shepard's method. Referred to as a vector of pixel values
    $\vec{S}$. 
    \item {\it Deconvolved reconstruction}: A model of the image 
      on the sky with the kernel deconvolved.
      Referred to as a vector of
    pixel values $\vec{F}$ 
    (never used directly in the final result).
    \item {\it Covariance-Regularized Reconstruction (CRR)}: The resampling resulting from our method.
      Referred to as a vector of pixel values $\vec{G}$.
    \item {\it Point spread function (PSF)}: In our usage, the PSF 
      will refer to the response to a delta-function on the sky
      of the output image from the analysis, either Shepard's image or the 
    CRR. In an ideal reconstruction, the PSF would be 
    identical to the kernel values at each pixel (using the
    delta-function location as the center of the kernel function).
\end{itemize}

This article is organized as follows. Section \ref{sec:resampling}
describes the resampling methods we consider here. Section 
\ref{sec:tests} presents a series of tests of this method for 
a simulation. Section \ref{sec:realdata} presents a demonstration
of the method for real data. Section \ref{sec:discussion} 
summarizes and discusses the results.

\section{Image Resampling Methods}
\label{sec:resampling}

\subsection{Shepard's method, the MaNGA DRP standard}
\label{sec:Shepard} 

A specialized version of Shepard's method is described in 
\citet{Sanchez2012} in the context of CALIFA and 
\citet{Law2016} in the context of MaNGA. The input data from 
the IFU survey consists of the flux intensity $f[i]$ and the variance 
of the error $N[i] = \sigma[i]^2 = \langle \Delta f[i]^2\rangle$ 
in each fiber. 
For a system with $N_{\rm fiber}$ fibers, with which we 
have taken $N_{\rm exp}$ exposures, for each wavelength channel
there are $N=N_\text{fiber}\times N_\text{exp}$ values, each 
corresponding to a different location on the sky. We will
refer below to each such observation as a fiber-exposure.
For the MaNGA DRP, the output grid is rectangular with a 
pixel size 0.5 arcsec per pixel, and we write the total number of
pixels as $M$.

The transformation from intensities at the irregularly located 
fiber locations to the Shepard's image is:
\begin{equation}
\label{eq:shepard}
S[j]=\sum_{i=1}^NW[j,i]f[i]
\end{equation}
where $\vec{S}$ is Shepard's image, and the $M\times N$ matrix $W[j,i]$ is the weight of each fiber 
location $i$ contributing to the output grid point $j$.
In Shepard's method, the weight function is a circularly 
symmetric  Gaussian that depends on the distance $r[i,j]$
between the fiber location $i$ and the grid point $j$:
\begin{equation}
\label{eq:shepardweights}
  W[j,i]=
  \frac{b[i]}{W_0[j]}\exp\left(-\frac{r[i,j]^2}{2\sigma_0^2}\right),
\end{equation}
for $r[i,j] < r_{\rm lim}$, and zero otherwise. 
$\sigma_0$ defines the width of the Gaussian function, and for the
MaNGA DRP is set to 0.7 arcsec. The MaNGA DRP takes
$r_\mathrm{lim}=1.6$ arcsec. The normalization parameter $W_0[j]$ is
defined as the sum of the $N$ weights for each output grid point $j$,
to guarantee the conservation of flux:
\begin{equation}
\label{eq:shepardnorm}
W_0[j]=\sum_{i}b[i]\exp\left(-\frac{r[i,j]^2}{2 \sigma_0^2}\right),
\end{equation}
over all pixels $i$ for which $r[i,j]< r_{\rm lim}$.
In Equations \ref{eq:shepardweights} and \ref{eq:shepardnorm}, 
$b[i]$ is zero if the inverse variance $N[i]^{-1}=0$ and is 
unity otherwise.

\subsection{Covariance-regularized reconstruction}\label{G_method}

Our method can be written in the same form as Shepard's method (Equation 
\ref{eq:shepard}),  but with a different choice of $W[j,i]$. The choice is 
motivated in the following manner, which explicitly designs
the weights so that the final reconstruction remains consistent
with the samples and so that its covariance matrix has small
off-diagonal entries. We refer to an image using our method
as a covariance-regularized reconstruction (CRR).

Consider fitting a linear model to reconstruct the observables $f[i]$.
Our model consists of a set of delta functions with fluxes $F[j]$,
distributed on a regular grid (the same grid we want to use for the
reconstruction). Each fiber-exposure $i$ observes this function convolved with
the kernel appropriate for that particular observation, $K_i(x,
y)$. In the case of MaNGA, the kernel is the convolution of the
atmospheric seeing, the telescope optical response, and the
fiber. This model for the observables can be written as
\begin{equation}
\label{eq:model}
m[i] = \sum
A[i,j]
F[j],
\end{equation}
where:
\begin{equation}
A[i,j] =  K_i(x_i-X_j, y_i-Y_j)
\end{equation}
is the value of the kernel at the separation between the model pixel
$j$ at $(X_j, Y_j)$ and the fiber-exposure $i$ at $(x_i, y_i)$. The
kernel is a known property of the observations based on the estimated
seeing at the time of each observation.  

We can fit for the model parameters $F[j]$ by minimizing
the $\chi^2$ error:
\begin{equation}
\label{eq:chi-square}
\chi^2=\sum_i\frac{\left(f[i]-m[i]\right)^2}{\sigma[i]^2}.
\end{equation}
Because the model is linear in $F[j]$, the solution can be written:
\begin{equation}
\label{eq:F}
\vec{F}=(\mathbf{A}^\mathrm{T}\mathbf{{N}}^{-1}\mathbf{A})^{-1}\mathbf{A}^\mathrm{T}\mathbf{N}^{-1}\vec{f},
\end{equation}
where $\mathbf{N}$ is the data covariance matrix, which in 
our case is diagonal with diagonal entries $\sigma[i]^2$.

Therefore, we can estimate $\vec{F}$, which is analogous to a
deconvolved image based on the fiber samples, because it is the sky
image before convolution with the kernel. As one expects for a
deconvolution, the error covariance for $\vec{F}$ is highly
non-diagonal --- there are strong correlations and anticorrelations in
the errors of neighboring pixels. The covariance matrix is
\begin{equation}
    \mathbf{C}=\left\langle \Delta \vec{F}\Delta \vec{F}^\mathrm{T} \right\rangle=(\mathbf{A}^\mathrm{T}\mathbf{N}^{-1}\mathbf{A})^{-1}.
\end{equation}
Using $\vec{F}$ for science is very undesirable because of 
these correlations, which would complicate any error analysis
but also lead to large fluctuations among the values of $\vec{F}$.

A common technique is to regularize the values of $\vec{F}$, 
either under a Tikhonov regularization, a maximum entropy criterion,
or something else (e.g. \citealt{warren03a}). However, here we 
take a different approach,
which is to regularize the covariance such that it is diagonal,
which turns out to be similar to reconvolving $\vec{F}$ to a 
resolution similar to the kernel. 

The covariance matrix can be whitened through a linear 
transformation of $\vec{F}$ that can be found by 
taking a square root of the inverse covariance matrix 
$\mathbf{C}^{-1}=\mathbf{A}^\mathrm{T}\mathbf{N}^{-1}\mathbf{A}$. 
For this symmetric and positive definite matrix, we can 
take its square root by finding its eigensystem:
\begin{equation}
\label{eq:QQ}
\mathbf{ C}^{-1}=\mathbf{ P}\mathbf{ D}\mathbf{ P}^{-1}=(\mathbf{ P}\mathbf{ D}^{\frac{1}{2}}\mathbf{ P}^{-1})(\mathbf{ P}\mathbf{ D}^{\frac{1}{2}}\mathbf{ P}^{-1})=\mathbf{ Q}\mathbf{ Q},
\end{equation}
where $\mathbf{D}$ here is a diagonal matrix of eigenvalues
and $\mathbf{P}$ is the matrix of eigenvectors. For 
$\mathbf{D}^{\frac{1}{2}}$, the positive root is always chosen.

Another path to finding the same matrix $\mathbf{Q}$ 
is through the singular value decomposition (SVD): 
\begin{equation}
\mathbf{ N}^{-\frac{1}{2}}\mathbf{ A}=\mathbf{U}\mathbf{\Sigma}\mathbf{V}^\mathrm{ T}.
\end{equation}
where $\mathbf{\Sigma}$ is diagonal, $\mathbf{V}$ is 
orthogonal ($\mathbf{V}\mathbf{V}^\mathrm{T} = \mathbf{1}$), and 
$\mathbf{U}$ is close to orthogonal
($\mathbf{U}\mathbf{U}^\mathrm{T}$ is diagonal with
ones for dimensions $i$ with $\Sigma_{i} \ne 0$
or zeros for dimensions $i$ with 
$\Sigma_i=0$).
In this case:
\begin{equation}
\vec{F} = \mathbf{V}\mathbf{\Sigma}^{-1}\mathbf{U}^T \mathbf{N}^{-\frac{1}{2}} \vec{f}
\end{equation}
if $\Sigma_i \ne 0$ for all $i$. We can also use the standard
Moore-Penrose inverse technique and set $\Sigma_i^{-1} = 0$
for $\Sigma_i = 0$, which allows us to handle truly degenerate
cases smoothly. We can show that
\begin{equation}
\mathbf{ C}^{-1}=\mathbf{ V}\mathbf{ \Sigma}\mathbf{ \Sigma}\mathbf{ V}^\mathrm{ T}
\end{equation}
and therefore,
\begin{equation}
\label{eq:Q}
\mathbf{ Q}=\mathbf{ V}\mathbf{ \Sigma}\mathbf{ V}^\mathrm{ T}.
\end{equation}
Thus, we can find $\mathbf{Q}$ without ever constructing the 
full covariance matrix or its inverse, or explicitly finding its
eigenvectors, which may be useful in cases when the covariance
matrix is ill-conditioned.

The linear transformation of $\vec{F}$ that we will use is not
quite $\mathbf{Q}$, because we want the transformation to 
conserve the total flux in $\vec{F}$.
We can achieve this goal by normalizing over each row 
to give the transformation matrix:
\begin{equation}
\label{eq:R}
R[i,j]= \frac{1}{\sum_j Q[i,j]} Q[i,j],
\end{equation}
$\mathbf{R}$ is then a linear transformation under which 
$\mathbf{C}$ can be diagonalized:
\begin{equation}
\label{eq:covariance_inv}
\mathbf{C}_G = \mathbf{ R}\mathbf{ C}\mathbf{ R}^\mathrm{T}
\end{equation}
with the entries of the diagonal matrix 
$\mathbf{C}_G$ given by
\begin{equation}
\label{eq:covariance_diagonal}
C_G[i,i]=\left(\sum_j Q[i,j]\right)^{-2}
\end{equation}
(In detail, the resulting covariance matrix is not precisely 
diagonal if any $\Sigma_i = 0$). 
This result means that if we define: 
\begin{eqnarray}
\label{G_formula}
\vec{G} 
&=& \mathbf{R}\vec{F}\cr 
&=& \mathbf{R} 
(\mathbf{A}^\mathrm{T}\mathbf{N}^{-1}\mathbf{A})^{-1}\mathbf{A}^\mathrm{T}
\mathbf{N}^{-1}\vec{f} \cr
&=& \mathbf{R} \mathbf{V} \mathbf{\Sigma}^{-1}  \mathbf{U}^\mathrm{T} \mathbf{N}^{-\frac{1}{2}} \vec{f} \cr
&\mathrel{\stackrel{\makebox[0pt]{\mbox{\normalfont\tiny def}}}{=}}& \mathbf{W} \vec{f}
\end{eqnarray}
Then $\mathbf{C}_G$ is the covariance matrix of $\mathbf{G}$. 
Multiplying by the matrix $\mathbf{R}$ turns out to be similar to 
a reconvolution of $\vec{F}$  with the kernel, although it is 
not strictly speaking a convolution.

$\vec{G}$ will be our reconstruction, which by design should have 
close to a  diagonal covariance matrix. We will show later that in 
the context of MaNGA it also has a sharper PSF than Shepard's 
method. Like  Shepard's method it is just 
a linear combination of the input fluxes so can be written in the 
same form as Equation \ref{eq:shepard}. 

There are some adjustments we will make to this method. The first
adjustment is that
we will apply a regularization term to handle singular values. 
This adjustment will make the procedure more numerically robust
but have almost no impact on the results. We can
think of this regularization term in terms of an adjustment to the
function we are minimizing in Equation \ref{eq:chi-square} with a 
quadratic term, the simplest version of Tikhonov regularization:
\begin{equation}
\label{eq:chi_square_reg}
\chi^2=\sum_i\frac{[f[i]-m[i])]^2}{\sigma[i]^2}+\lambda^2\sum_j F[j]^2
\end{equation}
The solution $\vec{F}$ is altered under this regularization to:
\begin{eqnarray}
\vec{F} &=& (\mathbf{A}^\mathrm{T}\mathbf{N}^{-1}\mathbf{A}+\lambda^2\mathbf{ I})^{-1}\mathbf{A}^\mathrm{T}\mathbf{N}^{-1}\vec{f} \cr
&=& 
\mathbf{V} \mathbf{\Sigma}_\ast^{-1}  \mathbf{U}^\mathrm{T} \mathbf{N}^{-\frac{1}{2}}
\vec{f}
\end{eqnarray}
where $\mathbf{\Sigma_{\ast}}$ is diagonal and:
\begin{equation}
\label{eq:sigma_ast}
\Sigma_{\ast, ii}^{-1} = \frac{\Sigma_{ii}}{\Sigma_{ii}^2 + \lambda^2}
\end{equation}
and the solution $\vec{G}$ becomes:
\begin{equation}
\label{G_formula_reg}
\vec{G} 
= \mathbf{R} 
\mathbf{V} \mathbf{\Sigma}_\ast^{-1}  \mathbf{U}^\mathrm{T }\mathbf{N}^{-\frac{1}{2}}
\vec{f}
\end{equation}

The second adjustment is that we will not use the actual flux noise
vector $\vec{\sigma}$ estimated from the data. If we did so, then
there would be a correlation between the flux in a fiber-exposure
(which determines the true noise vector) and the values in 
the corresponding column of $\mathbf{W}$.  This correlation will 
lead to a dependence of
the PSF in the CRR on the signal that will greatly
complicate the interpretation of the image. In addition, it 
is usually the case that the noise is estimated from the signal 
and using it as a weight in the fit will therefore be biased. 
For these reasons, in the calculation, we use 
$\mathbf{\tilde{N}}$, which equals unity where
$\mathbf{N}^{-1}$ is not zero, and zero where $\mathbf{N}^{-1}$ 
is zero.

A third adjustment we make is to slightly alter the kernel to remove
contributions for which the radius is larger than 4 arcsec. This
choice makes little difference in the final result but makes the
calculation considerably faster.

The fourth adjustment we make is to remove from the SVD calculation
any model pixels $j$ that are further than some distance $r_{\rm lim}$
(which we here set to 1.6 arcsec, the same as DRP's choice) from 
any fiber --- basically, any
pixels not well-constrained by the data --- which lowers the condition
number of $\mathbf{A}$ and makes the SVD more stable. 

The final method then can be written as follows:
\begin{equation}
\vec{G}=\mathbf{W}\vec{f},
\end{equation}
where:
\begin{equation}
\label{eq:reconstruction_w}
\mathbf{W}=\mathbf{R}\mathbf{V}\mathbf{\Sigma_\ast^{-1}}\mathbf{U}^\mathrm{T}\mathbf{\tilde{N}}^{-\frac{1}{2}}
\end{equation}
where $\mathbf{V}$, $\mathbf{U}$, and $\Sigma$ are from the SVD
of $\mathbf{A}$, $\Sigma_\ast$ is defined in Equation \ref{eq:sigma_ast}, 
and $\mathbf{R}$ is defined by Equations \ref{eq:Q} and \ref{eq:R}.
The covariance matrix of $\vec{G}$ is then:
\begin{equation}
\label{eq:covariance}
\mathbf{C}_G = \left\langle \mathbf{W} \Delta\vec{f} \Delta\vec{f}^\mathrm{T} \mathbf{W}^\mathrm{T} \right\rangle
=  \mathbf{W} \mathbf{N} \mathbf{W}^\mathrm{T}.
\end{equation}
Here, $\mathbf{N}$ is the diagonal variance matrix from the individual
fibers. This covariance matrix is not guaranteed to be diagonal, but 
because of how we have constructed the weights it
will prove to be much closer to diagonal than Shepard's method.

When operating on the resulting data cube for science, it is 
important to identify unreliable spaxels about which there is
little information provided by the fiber data.
The MaNGA pipelines use two masks, {\tt NOCOV} to indicate
that there is no information about the pixel, and 
{\tt LOWCOV} to indicate that there is a low amount of 
information about the pixel. 
These masks are based on the fiber-level maskbits, which
record the effect of hot pixels, cosmic ray hits, and other
effects that can make the data from a fiber unusable at 
some or all wavelengths. Our approach to defining these
masks for the data cube spaxels is different from the standard 
MaNGA pipeline. The MaNGA pipeline counts the total contribution
of fibers to each pixel, based on the ${\mathbf W}$ matrix for
Shepard's method, to identify poorly constrained spaxels. 
We use a different approach, which is to use the diagonal 
elements of the covariance matrix when we assume constant,
unit noise in each fiber that is not masked:
\begin{equation}
\mathbf{\tilde{C}} = {\mathbf W} \mathbf{\tilde{N}} {\mathbf W}^T.
\end{equation}
This covariance matrix is not the same as the actual
covariance ${\mathbf C}_G$. We set spaxels with a variance
more than twice the median variance to {\tt LOWCOV}, so that 
they may be ignored in scientific analysis. This procedure
appropriately masks spaxels too near the edge of the fiber
bundle or that are affected by bad fibers.

There are several free parameters in the CRR method as applied to
MaNGA, which we now discuss. First, there is the
shape of the kernel, which is determined by our estimate of the
observing conditions at each exposure. Second, there are the
conditions for dropping edge pixels. Third, there is the
regularization parameter $\lambda$. Fourth, there is the pixel scale
of CRR. The first three parameters prove to have no
significant affect on the results for MaNGA, as we show later in this
paper. We will explore below the effect of the reconstruction pixel
scale.

Some general aspects of the method, compared to Shepard's 
method, are worth noting before describing those tests. Although
our method is also just a linear combination of the fluxes, the 
weights are not determined by a stationary function as they are
for Shepard's method. They are also not restricted to be 
nonnegative. These properties, particularly the latter, are 
essential to reducing the  off-diagonal covariances and producing 
a PSF close to the kernel resolution.

Although we motivated the method based 
on the model fit to the parameters $\vec{F}$ expressed in Equation 
\ref{eq:chi-square} and  the diagonalization of their covariance 
matrix $\mathbf{C}$, we never need to explicitly determine 
either $\vec{F}$ or $\mathbf{C}$. We will nevertheless calculate 
these quantities below in order to demonstrate their properties.

Our approach is mathematically identical to the approach 
proposed by \cite{Bolton2010} in the different context of 
spectroscopic extraction. Our notation differs somewhat
from theirs. Specifically, their $\vec{p}$ is our $\vec{f}$,
their $\vec{f}$ is our $\vec{F}$, and their $\tilde{f}$ is
our $\vec{G}$.

\section{Tests on Simulated Data}
\label{sec:tests}

In this section, we use simulations to characterize the 
performance of Shepard's method, which is what the MaNGA DRP uses,
and our CRR method. 

\subsection{General information}\label{general_information}

\begin{figure*}
\begin{center}
\includegraphics[width=1\textwidth]{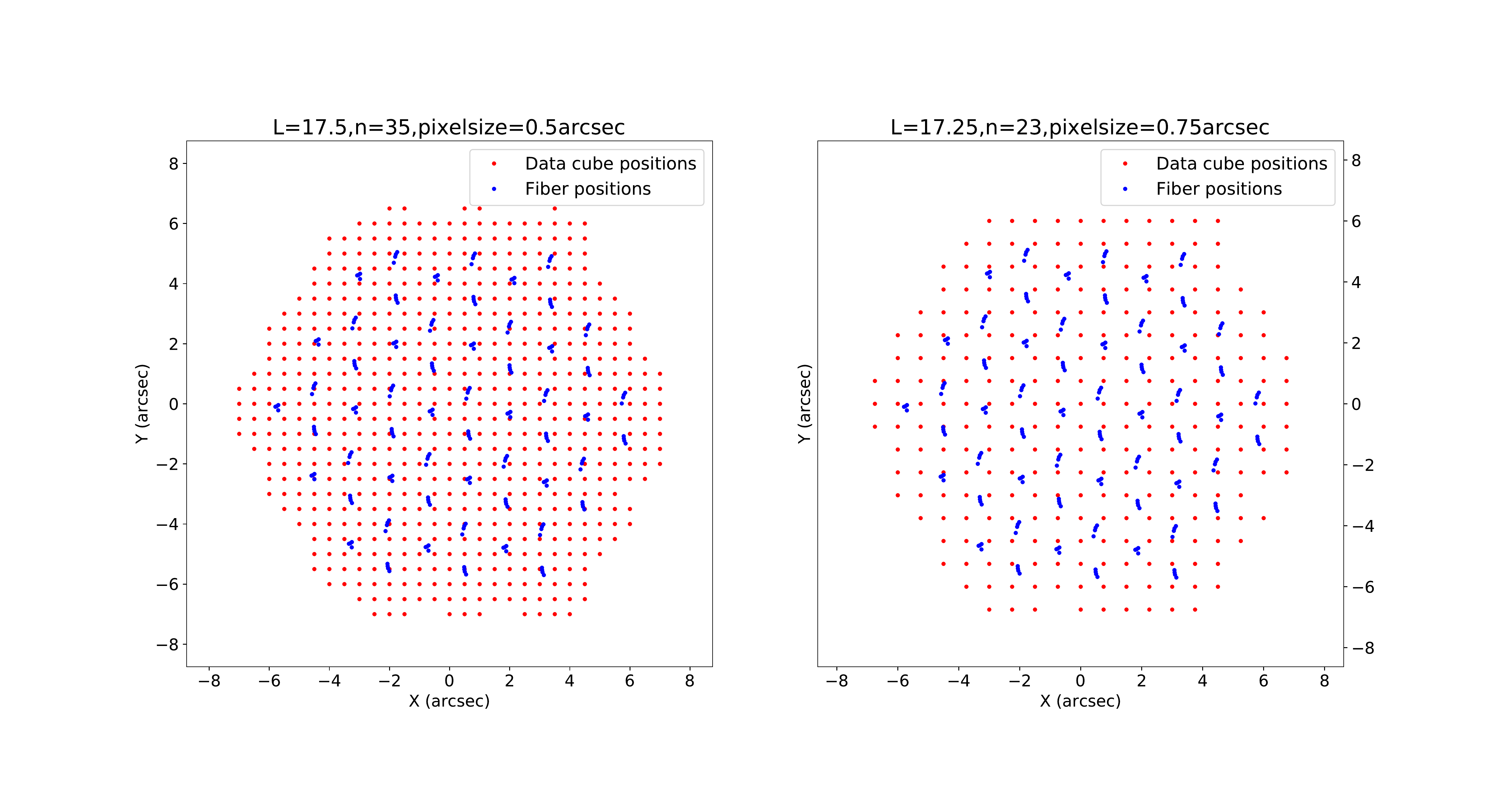}
\vspace*{-15mm}
\caption{Configuration of plate-IFU 8720-1901 with two different 
pixel scales, for wavelength  $\lambda=5500\angstrom$. Blue 
points are IFU fiber locations, red points are  pixels in the 
reconstruction. The left panel shows pixel scale 0.5 
arcsec/pixel, and the right panel shows pixel scale 0.75 arcsec/pixel.
\label{de_8720-1901_configuration}}
\end{center}
\end{figure*}

MaNGA uses a hexagonal pattern of fibers to detect flux as a function
of position on the sky. The hexagonal pattern is dithered in position
between different exposures, and varies on the sky with wavelength.
Our goal is to resample the fluxes onto a rectilinear grid.  In Figure
\ref{de_8720-1901_configuration}, the blue points show the fiber
locations at $\lambda=5500$\angstrom\ for the plate-IFU 8720-1901,
shown as a function of $X$ and $Y$ position in the focal plane
(corresponding to RA and Dec). The red points show the pixels we are
using in the reconstruction. As explained above, we exclude pixels
when their distance to all the fibers is larger than 1.6 arcsec.  In
this figure, we show both 0.5 arcsec/pixel and 0.75 arcsec/pixel
scales. We will examine the effect of difference choices of pixel
scales on our reconstruction later.

We will simulate observations for a point source to test
the methods. Since the methods we test are linear, the point 
source responses at  different locations can fully characterize 
the performance. The flux for each fiber is sampled from 
the kernel function $K$ as the response of a point source at some 
location (in most cases, we use $X=Y=0$).  

\begin{figure}
\begin{center}
\includegraphics[trim=2cm 4cm 0cm 1cm,width=0.5\textwidth]{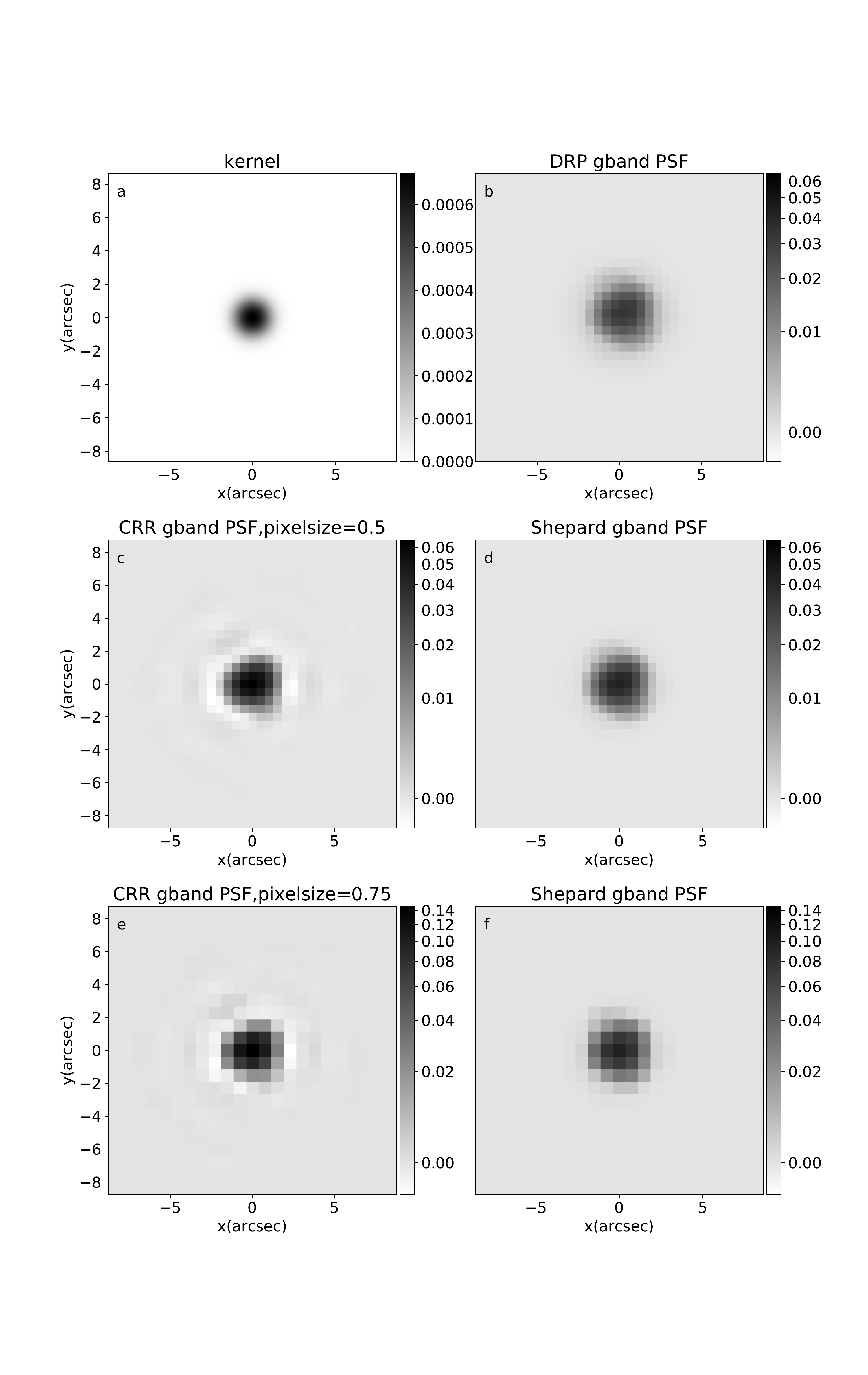}
\caption{Data cube slices from simulated observations of
a point source generated for plate-IFU 8720-1901. The flux is 
averaged over the $g$-band, and we assume seeing at 
5500$\angstrom$ is around 1.19 arcsec. 
Panel (a): the kernel for wavelength $\lambda=5500\angstrom$. 
Panel (b): the MaNGA DRP implementation of Shepard's 
method (0.5 arcsec pixels). 
Panel (c): reconstruction using the CRR
method (0.5 arcsec pixels). Panel (d): our implementation of Shepard's method (0.5 arcsec pixels). Panel (e): reconstruction using the CRR
method (0.75 arcsec pixels). Panel (f): our implementation of Shepard's method (0.75 arcsec pixels). The image scaling is linear for Panel (a),
but is arcsinh-scaled for Panels (b)--(f) to show the behavior in 
the PSF wings.}
\label{de_8720-1901_PSF_distribution_gband}
\end{center}
\end{figure}

Panel (a) of Figure 
\ref{de_8720-1901_PSF_distribution_gband} shows our adopted kernel for
wavelength $\lambda=5500\angstrom$. 
We use a double Gaussian function to simulate the atmospheric 
seeing, and convolve with the fiber profile (a 2 arcsec top-hat) 
to generate the kernel function. The normalized double Gaussian is 
defined  by the standard deviation $\sigma_1$ for the inner Gaussian, 
the ratio $\sigma_2/\sigma_1$ of the outer to inner Gaussian standard
deviation, and the ratio $A_2/A_1$ of the central values of the 
outer and inner Gaussians. We use a typical pair of ratios that 
approximates the inner parts of a Moffat-like atmospheric PSF (\citealt{Law2015}; Jim  Gunn, private communication):
\begin{eqnarray}
\frac{\sigma_2}{\sigma_1} &=& 2\cr
\frac{A_2}{A_1} &=& \frac{1}{9}
\end{eqnarray}
The FWHM of the resulting model PSF is:
\begin{equation}
\text{FWHM}=1.05\text{~FWHM}_1=2.473\sigma_1 
\end{equation}
The MaNGA data reduction reports the FWHM of atmospheric seeing 
for each exposure at the guider wavelength 
$\lambda_0=5400\angstrom$. For other wavelengths, we will assume
the seeing varies as $\lambda^{-1/5}$ \citep{Yan2016}.

For each simulation, the fiber locations 
are provided by a MaNGA observation with several exposures, for which positions accounting for a variety of observational effects (bundle metrology, dithering, chromatic and field differential refraction, etc.) are stored in the MaNGA data products, for example those shown in Figure \ref{de_8720-1901_configuration}. 
Normally the number of MaNGA exposures ranges from 6--21 in order 
to achieve uniformity and the required $S/N$ ratio.

\begin{figure}
\begin{center}
\includegraphics[width=0.5\textwidth]{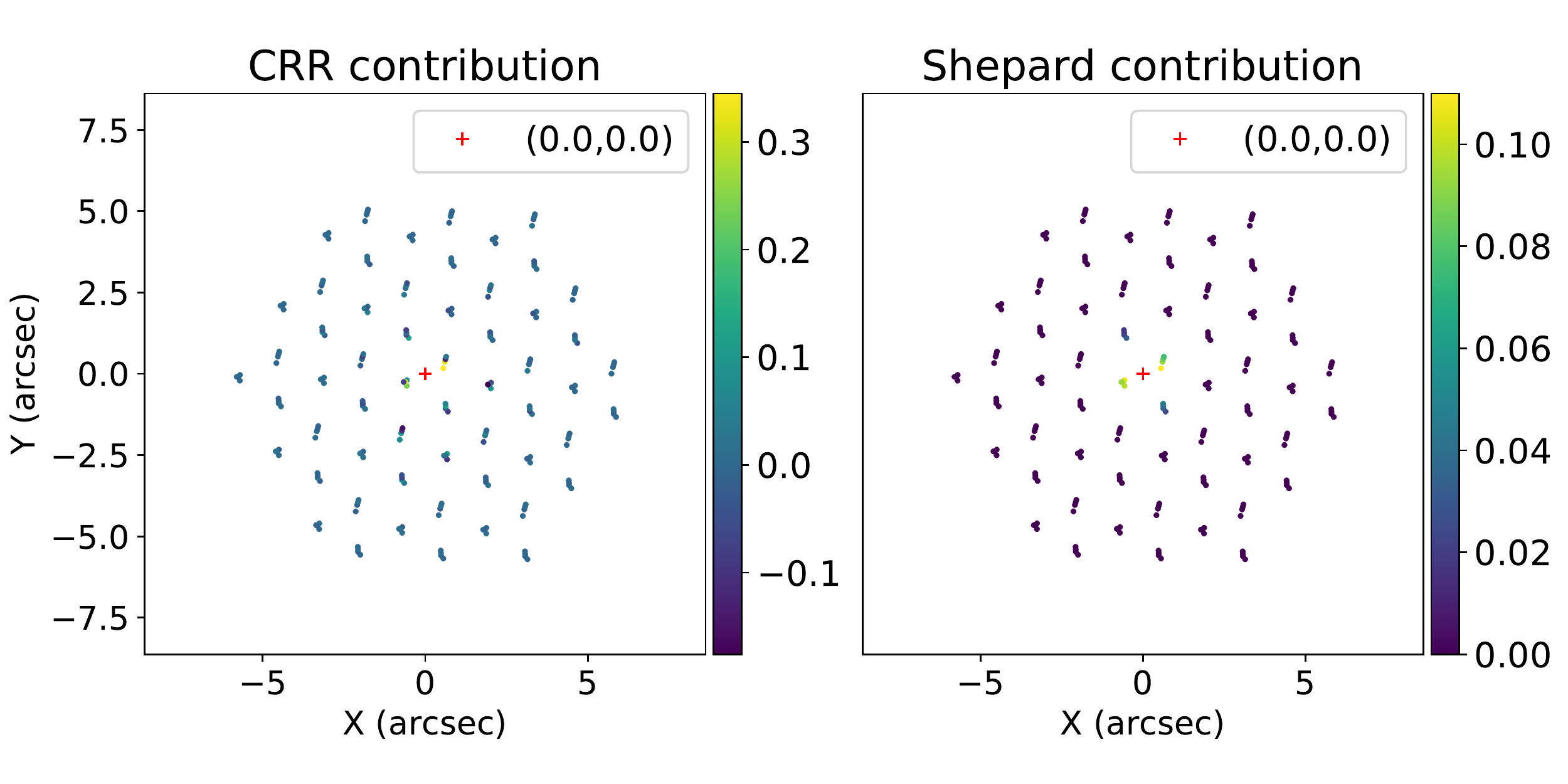}
\caption{The contribution of all fibers on the pixel at (0,0) for plate-IFU=8720-1901, which is a row of weights matrix $\mathbf{W}$. $\lambda=5500\angstrom$. Left panel: our CRR image. Right panel: Shepard's method. The red cross is the central pixel evaluated in the graph.}
\label{de_8720-1901_contribution}
\end{center}
\end{figure}

Each pixel in the output image is a weighted sum of the values
sampled for each fiber-exposure. Figure 
\ref{de_8720-1901_contribution} shows an example for the central pixel. 
For Shepard's weights (right panel), the weights for each fiber-exposure 
are just a decreasing function of distance from the pixel. For 
the CRR, the weights are both positive and negative and have an
oscillatory nature, similar to that found in sinc-interpolation
methods.

Whereas the fluxes sampled from the kernel are noiseless, the 
fluxes of the actual observations are not. The main sources of noise
are Poisson noise in the number of electrons due to the object,
sky, and dark current, plus the read noise from the amplifiers. 
For the bulk of locations and wavelengths in MaNGA, the noise is 
object-dominated. When we include simulated noise in our tests,
we concentrate on this regime, so that the noise is 
proportional to the square root of the flux. We characterize
the signal-to-noise ratio of the simulations based on that
of the fiber with the maximum simulated (noiseless) flux 
$f_{\rm max}$. For a chosen signal-to-noise ratio we then 
define a scale factor $s$ converting flux to number of 
photons $N = fs$, so that:
\begin{equation}
S/N=\frac{f_\mathrm{max}\cdot s}{\sqrt{f_\mathrm{max}\cdot s}}=\sqrt{f_\mathrm{max}\cdot s}
\end{equation}
Then for each fiber we apply Poisson noise based on the resulting
$N$ for each fiber.

\subsection{Nominal case}\label{test_nominal}

We start our tests with the nominal case at 0.5 arcsec/pixel, 
in order to compare directly at the MaNGA DRP pixel scale.
Figure \ref{de_8720-1901_PSF_distribution_gband} shows the 
response of a point source in the center, averaged
over the $g$-band wavelength range. Panel (b), (c), and (d)
show, respectively, MaNGA's DRP datacube (which uses
Shepard's method), the CRR image, and 
our implementation of Shepard's image, all
at 0.5 arcsec/pixel. Panel (e) and (f) show the CRR image and 
Shepard's image at 0.75 arcsec/pixel. As we said previously, 
panel (b) and (d) are almost the same because they use the same 
method. We show the DRP result in (b) to confirm that we are 
analyzing the data consistently. 

We compare the performance of the methods quantitatively
by measuring the FWHM and also a pseudo-Strehl ratio for 
the resulting PSF. We 
define the pseudo-Strehl ratio as the ratio of the peak image 
data cube slice intensity compared to the center of the 
kernel-convolved point source image. This is the ratio 
of the peak intensity to the largest possible given the
resolution of the instrument. We show here a single typical 
wavelength slice at $\lambda=5500\angstrom$ with seeing 
FWHM around 1.19 arcsec, but we find  similar results 
throughout the whole spectrum and various plate-IFUs.

\begin{figure}
\begin{center}
\includegraphics[width=0.5\textwidth]{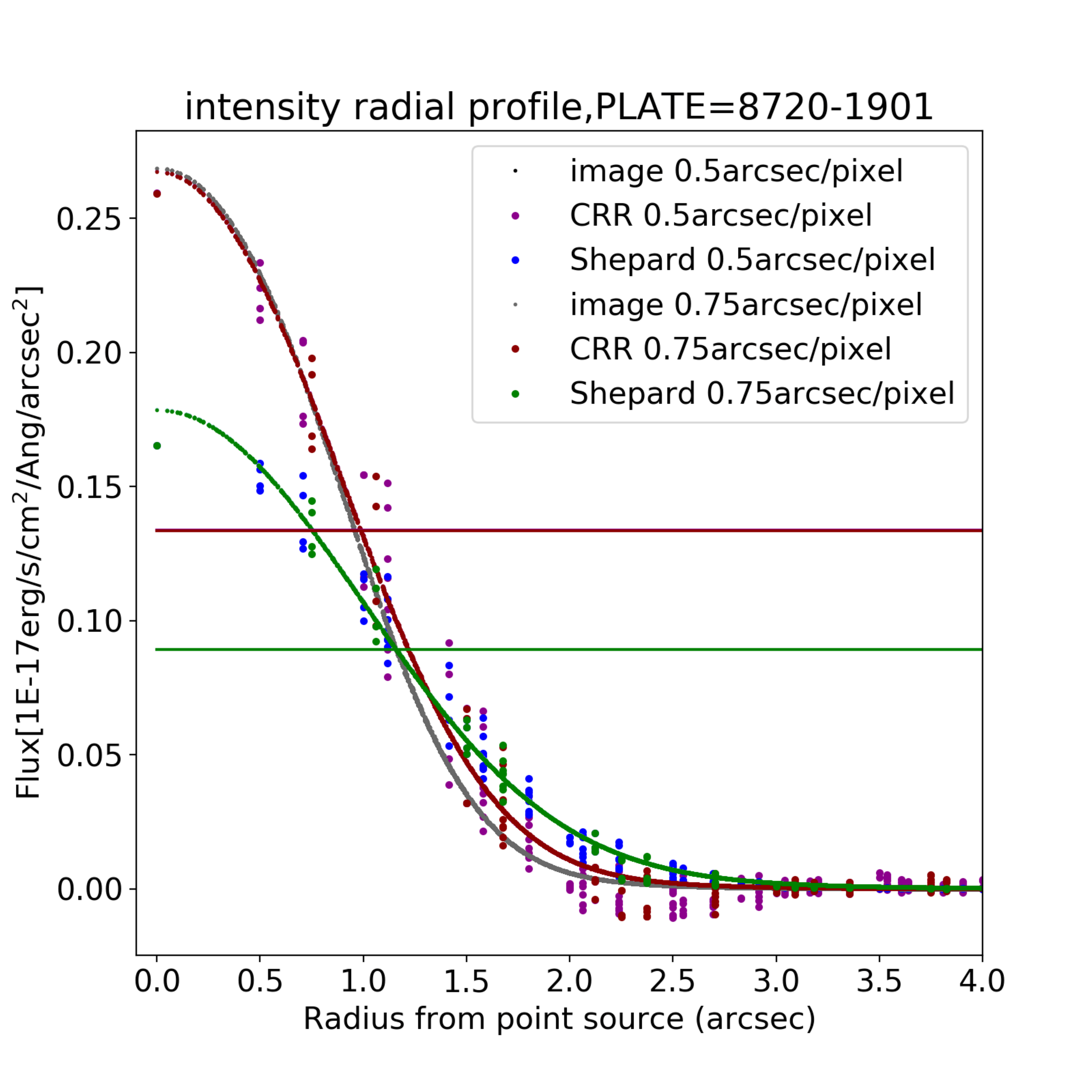}
\caption{Profile of the PSF in the output images from the simulated 
data in Figure \ref{de_8720-1901_PSF_distribution_gband}. Points
are flux values per pixel throughout the image at 
0.5 arcsec/pixel and 0.75 arcsec/pixel, for wavelength $\lambda=5500\angstrom$. 
The horizontal lines indicate the half-maximum for each radial profile
fit.}
\label{dradial_8720-1901_numerical}
\end{center}
\end{figure}

Figure \ref{dradial_8720-1901_numerical} shows the profile 
of the intensity of the resulting images. 
To determine a FWHM reliably, we fit a model to the 
pixel values. In particular, we use the kernel function
$K$ as our model, which is parametrized by 
a choice of seeing. We use least-squares to 
find the best-fitting seeing value, and infer the 
FWHM and the pseudo-Strehl ratio from the corresponding kernel. 
In analyzing the MaNGA DRP results, \citet{Law2016} used 
a Gaussian fit instead; we find that this is adequate for 
a broad PSF but not for the CRR image PSF.
The results for plate-IFU 8720-1901, including 
the FWHM and the Strehl ratio for the kernel, for Shepard's
method, and for the CRR method, are listed in Table 
\ref{table_analysis}.

\begin{center}
\begin{table}
\caption {Sharpness measures of PSF.}\label{table_analysis}
\begin{tabular}{c|c|c|c|c}
\hline
\backslashbox{analysis}{Item}& \shortstack{pixel scale\\ per pixel} & Kernel & Shepard's&CRR\\
\hline
FWHM(arcsec) & 0.5 & 1.905& 2.316 &1.973\\
Strehl ratio & 0.5 & 1 &0.664&0.996\\
\hline
FWHM(arcsec) & 0.75 & 1.905&2.317&1.973\\
Strehl ratio & 0.75 & 1 &0.664&0.996\\
\hline
\end{tabular}
\end{table}
\end{center}

For consistency, given a pair of Strehl ratios or FWHM values 
$v_1$ and $v_2$ for the Shepard and CRR images, we define the 
improvement in this quantity as $\frac{2|v_2-v_1|}{v_2+v_1}$.
The radial profile shows that there is a 16.0\% percent 
improvement in the FWHM between Shepard's method and CRR, 
which  turns out to be a typical level of improvement. 
Denothing the kernel FWHM as $v_0$, the maximum possible improvement 
of the FWHM would be $\frac{2(v_2-v_0)}{v_2+v_0}=19.5\%$ at most.
The pseudo-Strehl ratio increase is around 40\%. Our 
results  are very close to the 
best possible performance without performing a deconvolution
(i.e. very close to the kernel).
We note in passing that we were surprised
that the kernel had a FWHM slightly less than 2 arcsec, given
that the  fiber top-hat FWHM is precisely 2 arcsec; however, 
this result is correct --- slightly blurring the top-hat with 
a Gaussian slightly reduces the FWHM.\\

Next we compare the covariance between pixels for both methods. 
The covariance between pixels of the final image is 
expressed in Equation \ref{eq:covariance}, for both our method or 
Shepard's method, depending only on which weights $\mathbf{W}$ are 
used. In order to factor out the magnitude of the covariance
along the diagonal, we will examine the correlation matrix between
pixels, defined in the usual way as
$\rho_{jk}=C_{jk}/\sqrt{C_{jj}C_{kk}}$. 

\begin{figure}
\begin{center}
\includegraphics[width=0.5\textwidth]{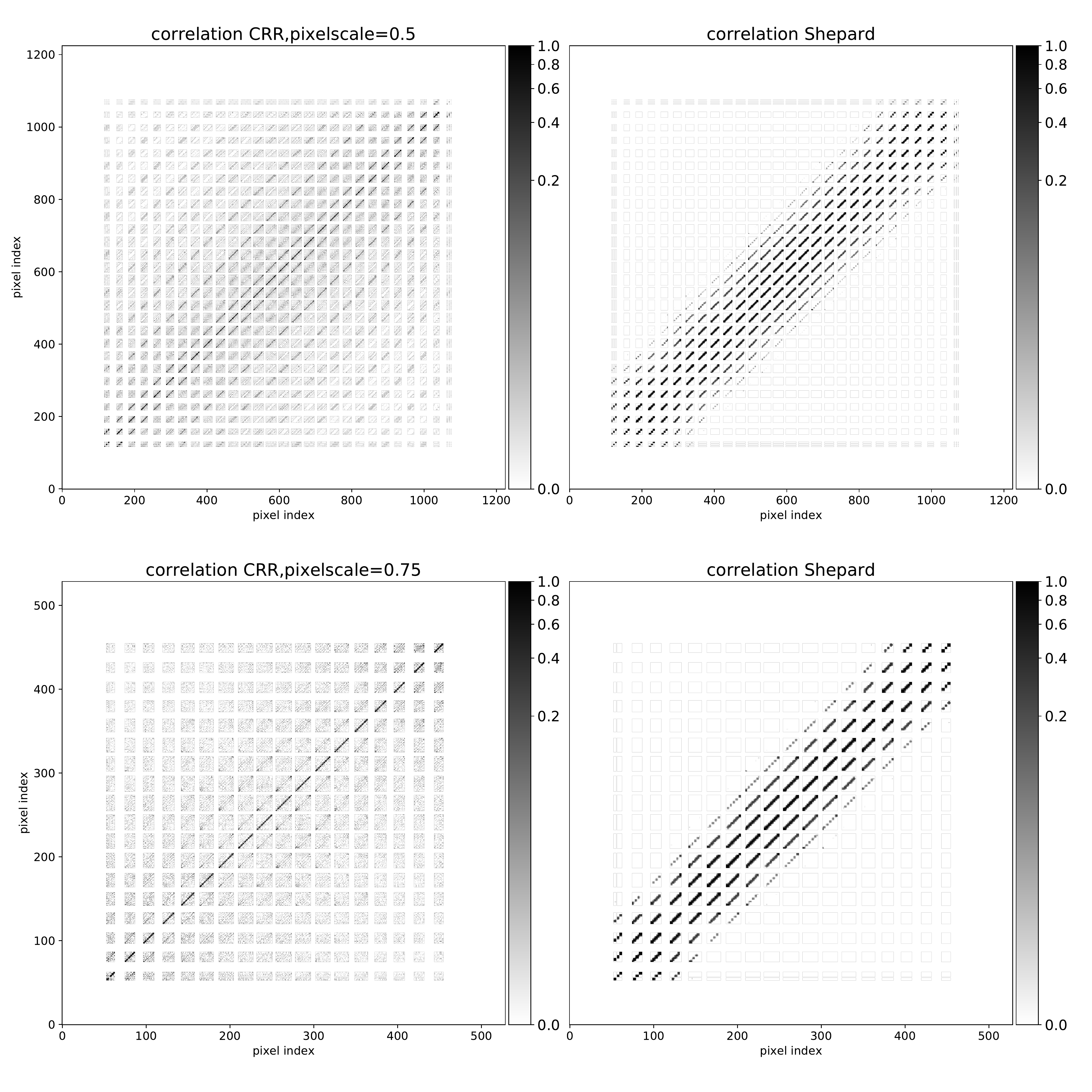}
\caption{The correlation matrix of CRR and Shepard's result 
for plate-IFU 8720-1901, both shown at 0.5 arcsec/pixel and 
0.75 arcsec/pixel, for wavelength $\lambda=5500\angstrom$. The 
image scaling is arcsinh to better see the small off-diagonal elements.
Left panel: CRR result. The correlation coefficients between values 
separated by two or more pixels are typically of the order 
$10^{-2}$ or less. Right panel: Shepard's result.}
\label{de_8720-1901_cov}
\end{center}
\end{figure}

Figure \ref{de_8720-1901_cov} shows the resulting correlation
matrix between  pixels. We omit masked pixels, which are mostly 
at the corners. The image scaling is arcsinh to best examine
the off-diagonal components.
Shepard's output image has a broad covariance matrix, 
while our method is nearly diagonal, as expected from the previous
section.   We show pixel scales of 0.5 and 0.75
arcsec/pixel, and for the CRR covariances the difference in the 
level of correlation between these cases is clear.  We consider 
the choice of pixel size and its effect on 
covariance more quantitatively in the  next subsection.

\subsection{Choice of pixel size}

Here we discuss the choice of pixel size. The primary 
considerations are the sharpness of the resulting PSF
(as quantified by FWHM and pseudo-Strehl ratio), the 
covariance between pixels, and the sampling of the 
image. The results shown in this section for 
plate-IFU 8720-1901 and $\lambda = 5500$ 
\AA\ are representative of what we find at other wavelengths
and in other plate-IFUs. 

The sharpness of the PSF differs very little between
the two pixel scales, as Table \ref{table_analysis} demonstrates.

\begin{figure}
\begin{center}
\includegraphics[width=0.5\textwidth]{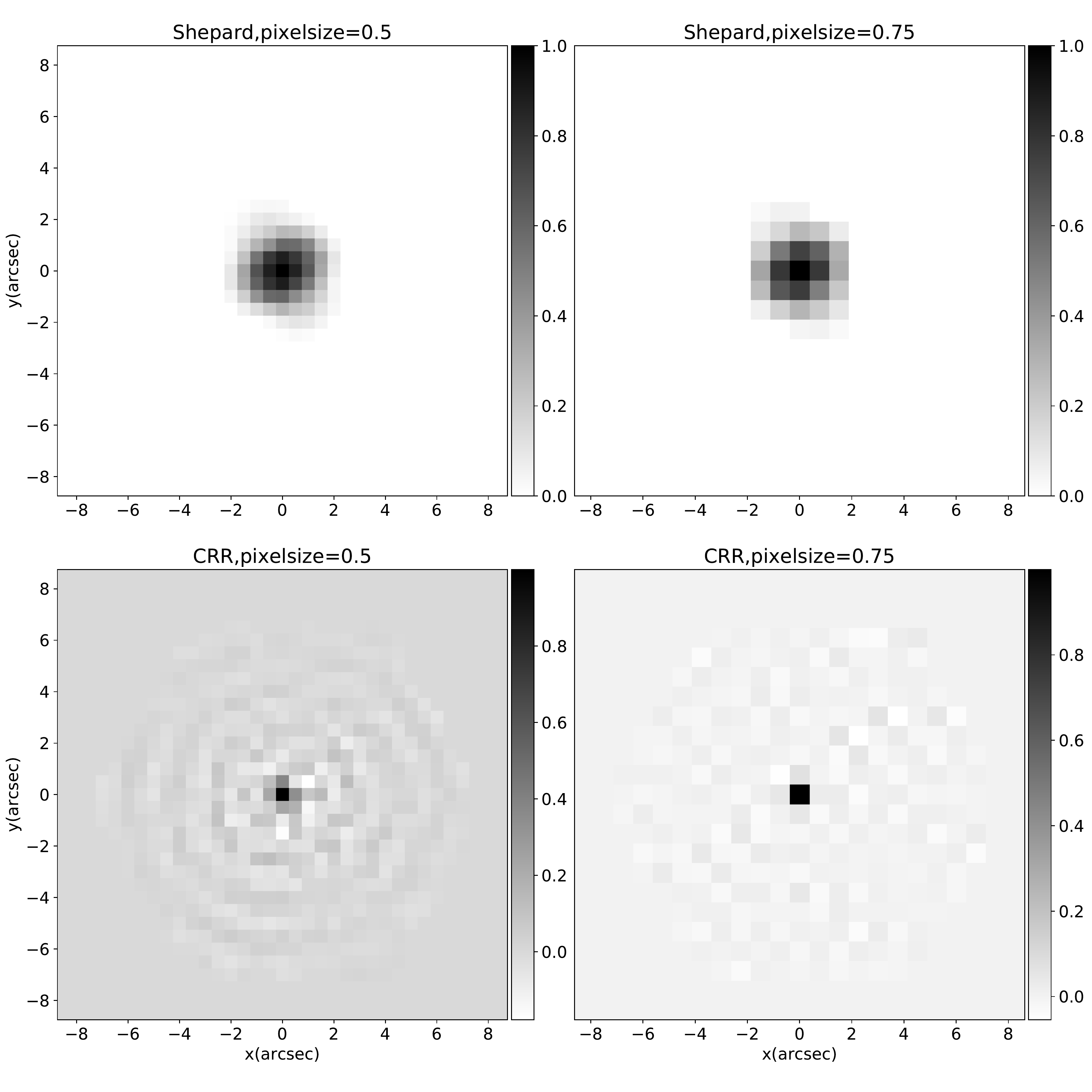}
\caption{The correlation between the central pixel flux and other pixels
for images produced from plate-IFU 8720-1901, at wavelength
$\lambda=5500\angstrom$. Upper panels: 
Shepard's method at 0.5 arcsec/pixel and 0.75 arcsec/pixel. 
Lower panel: our method at 0.5 arcsec/pixel and 0.75 arcsec/pixel. }
\label{de_8720-1901_corr}
\end{center}
\end{figure}

The correlation coefficients do tend to depend on 
pixel scale, as shown in Table \ref{table_corr},  based on 
the central pixel of plate-IFU 8720-1901 at 5500 \AA. In this table,
we list the quadratic means of the correlation coefficients between
the central pixel and the pixels which are separated from it
by 1 pixel or 2 pixels. 
We  examine the correlations more fully in Figure 
\ref{de_8720-1901_corr}, which shows the full correlation between the central pixel and all others for both methods with the two pixel
scales. Each image corresponds to a row of the correlation
matrix. The CRR image outperforms the Shepard's image in either case. The CRR image has some ringing in its correlation matrix that is greatly reduced at a pixel scale 
of 0.75 arcsec/pixel. Statistical independence is desirable so that
the fluxes in the cube can be used to fit models and to 
propagate errors without tracking a broad covariance matrix, 
which is complex and burdensome.
\begin{center}
\begin{table}
\caption {Correlation coefficient between
  pixels with different separations (quadratic mean of values between
  the central pixel and its surrounding pixels), for Shepard's method
  and CRR.}\label{table_corr}
\begin{tabular}{c|c|c|c}
\hline
\backslashbox{method}{separation}& \shortstack{pixel scale\\ arcsec/pixel} & 1 pixel & 2 pixels\\
\hline
 Shepard's & 0.5 & 0.876& 0.703 \\
  Shepard's & 0.75 & 0.767 & 0.465 \\
  CRR & 0.5 & 0.303 & 0.094\\
  CRR & 0.75 & 0.050 &0.024\\
\hline
\end{tabular}
\end{table}
\end{center}
The final major consideration for pixel size is whether the pixels
provide sufficient sampling of the image. At 0.75 arcsec/pixel, 
the FWHM of the images are about 2.5 pixels, which is above the 
usual rough guide for critical sampling. Although for unusual 
PSFs of the sort produced by this instrument there may yet be 
some poorly sampled power at 0.75 arcsec/pixel, 
sampling more densely leads to an increase in covariance
between pixels, as shown above.

These considerations of sampling and covariance lead us to 
0.75 arcsec/pixel as the best choice for our application.
This pixel scale provides decent sampling without inducing 
excess correlations between pixels or ringing behavior.

\subsection{PSF as a function of source position}\label{test_location}

\begin{figure}
\begin{center}
\includegraphics[trim=2cm 0cm 0cm 0cm, width=0.5\textwidth]{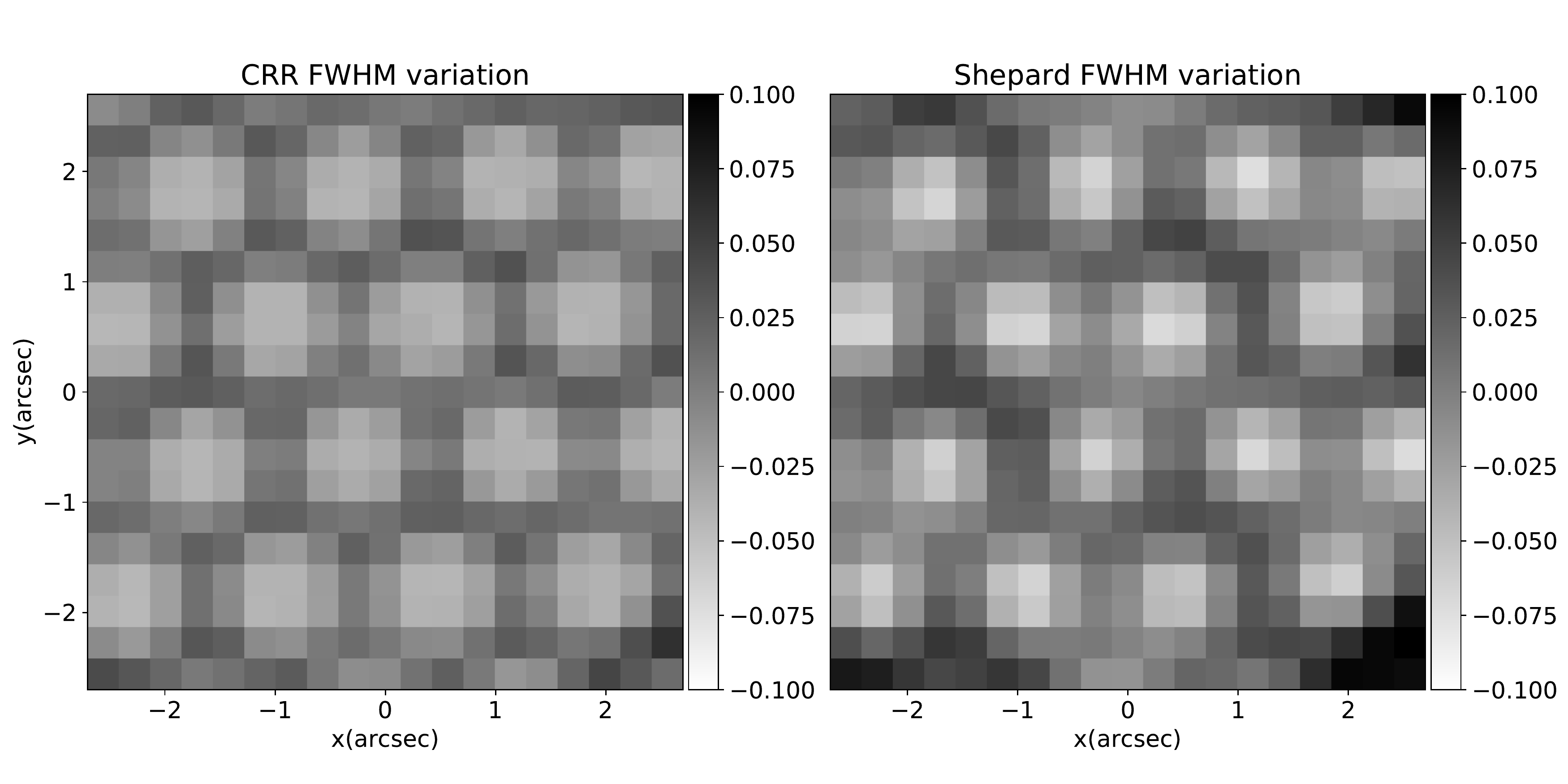}
\caption{Fractional FWHM variation as a function of point source location,
for plate-IFU=8720-1901 at wavelength $\lambda = 5500$ \AA,
Left panel: CRR image, using 0.75 arcsec/pixel. Right panel: 
Shepard's image, using 0.5 arcsec/pixel to simulate 
the MaNGA DRP performance.}
\label{dlocation_8720-1901_FWHM}
\end{center}
\end{figure}

\begin{figure}
\begin{center}
\includegraphics[trim=2cm 0cm 0cm 0cm, width=0.5\textwidth]{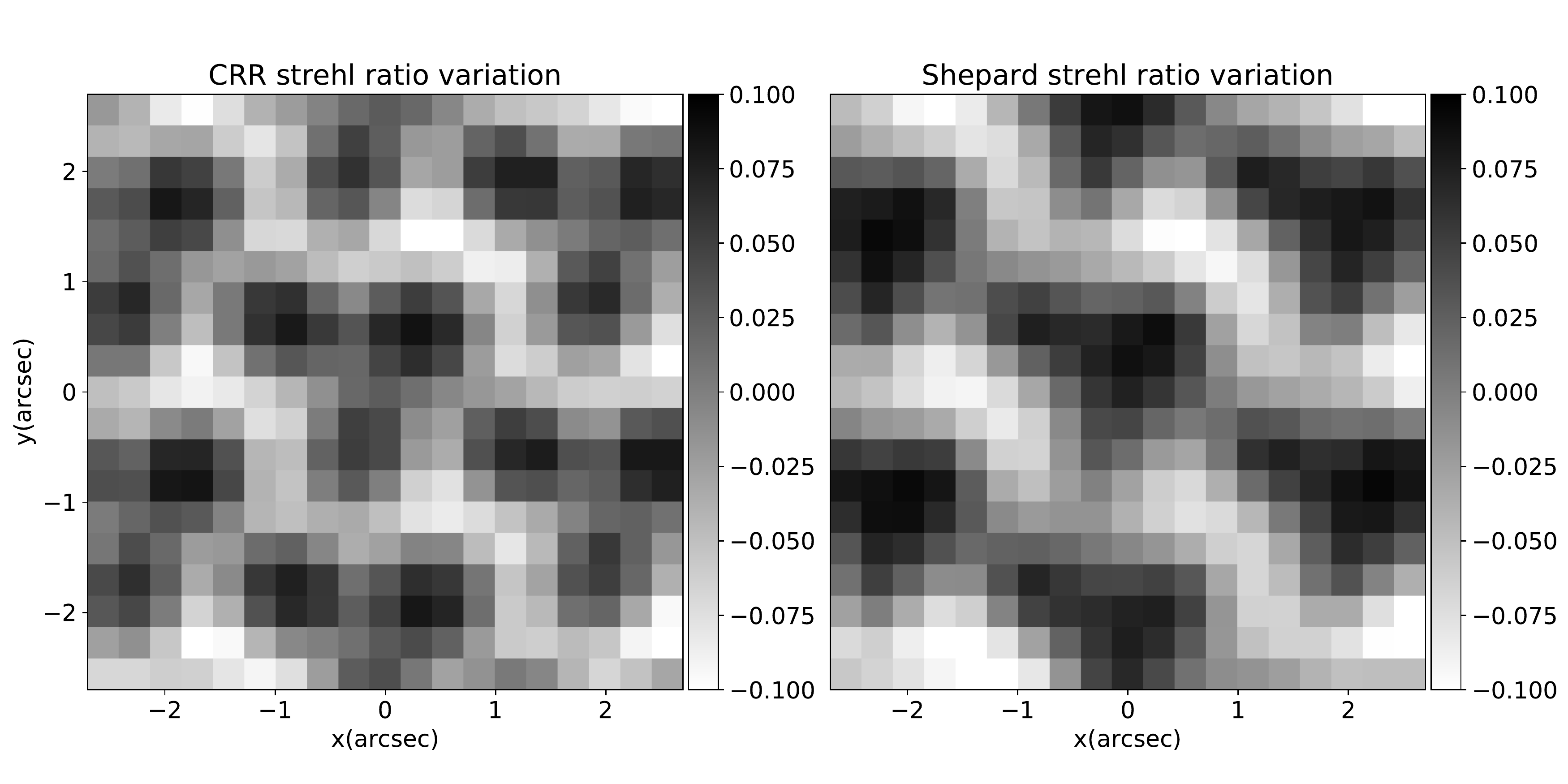}
\caption{Similar to Figure \ref{dlocation_8720-1901_FWHM}, for
the fractional Strehl ratio variation.}
\label{dlocation_8720-1901_strehl}
\end{center}
\end{figure}

The PSF response for both CRR and Shepard's is sensitive 
to the relative position on the sky of the point source and the fibers.
We therefore need to characterize the PSF response across the face of 
IFU and verify that we satisfy MaNGA's requirement that the PSF FWHM
vary by less than 10\% across the IFU.
In Figure \ref{dlocation_8720-1901_FWHM} we consider the 
FWHM homogeneity and in 
Figure \ref{dlocation_8720-1901_strehl} we consider the 
pseudo-Strehl ratio homogeneity.
Below, we also discuss the  homogeneity of the 
axis ratio $b/a$ of the PSF.

Each figure uses a grid of  locations of the point source, 
using the 0.75  arcsec/pixel CRR  and the 0.5 arcsec/pixel 
Shepard's method results (to match MaNGA's implementation). 
We quantify the variation of each quantity
relative to its median: 
\begin{equation}
\delta\sigma=\frac{\sigma-\sigma_0}{\sigma_0}.
\end{equation}
Here, $\sigma$ is the FWHM or 
the pseudo-Strehl ratio, and $\sigma_0$ is its median value.
As noted above, the CRR image is consistently higher
resolution than the Shepard's image for point sources 
regardless of where they are relative to the fibers. 
In addition, in this case the CRR image also shows 
less fractional variation in  resolution than Shepard's 
method, especially in the FWHM. In the example 
plate-IFU 8720-1901, the fractional standard 
deviation of the FWHM is 0.033 for Shepard's method, compared 
to 0.023 for our method. Meanwhile, the fractional standard 
deviation of the pseudo-Strehl ratio is 0.056 in Shepard's method,
compared to 0.048 in our method. Thus, the CRR
image slightly outperforms Shepard's image in terms of 
PSF homogeneity for this fiber bundle. 
This comparison depends on the location of the fiber-exposures; for
the other fiber bundles we have tested, the CRR images always give 
similar or better homogeneity than Shepard's images for the 
FWHM and pseudo-Strehl ratio.

However, for the PSF roundness we find that Shepard's method 
outperforms CRR. We quantify the roundness with the axis ratio
$b/a$ of the PSF. For each point source location within a radial distance to the center equalling to the FWHM, we fit the PSF with a 2D Gaussian. The parameters of this fit yield the 
minor-to-major axis ratio. Following \citep{Law2015}, 
we quantify the performance with $(b/a)_{99}$, which 
is the axis ratio for which 99\% of the point source 
locations yield a PSF with $b/a > (b/a)_{99}$. 
This quantity is a conservative lower limit on the PSF 
roundness and its variation. In order to compare more
directly to \cite{Law2015}, we implemented our test with 
constant seeing. In our analysis, Shepard's 
image has $(b/a)_{99} = 0.91$, whereas the CRR image has 
$(b/a)_{99} = 0.85$. 

We were unable to reproduce the 
results of \citep{Law2015} for Shepard's method, who 
found a substantially higher value of $(b/a)_{99} = 0.96$
using a very similar methodology to ours. This is partly because the coordinates we adopt is from an actual observational fiber bundle. If we use a three-point dithered hexagonal grid with seeing FWHM=1.4 arcsec to do the simulation, we got $(b/a)_{99} = 0.942$ with Shepard's reconstruction method. The homogeneity statistics parameters are consistently better than with the real observational coordinates. But we also found Shepard's method outperforms our result in the homogeneity.


\subsection{Adding noise}

The simulated data we test on above is noiseless. However,
we need to test our performance in the presence of realistic
noise. The sources of noise in MaNGA are outlined in Section 
\ref{general_information}, but as noted there we will only
consider the effect of Poisson noise due to the object signal,
which is the most problematic (increasing with the signal 
rather than staying constant) and which is usually dominant.

\begin{figure}
\begin{center}
\includegraphics[trim=2cm 3cm 3cm 0.1cm, width=0.5\textwidth]{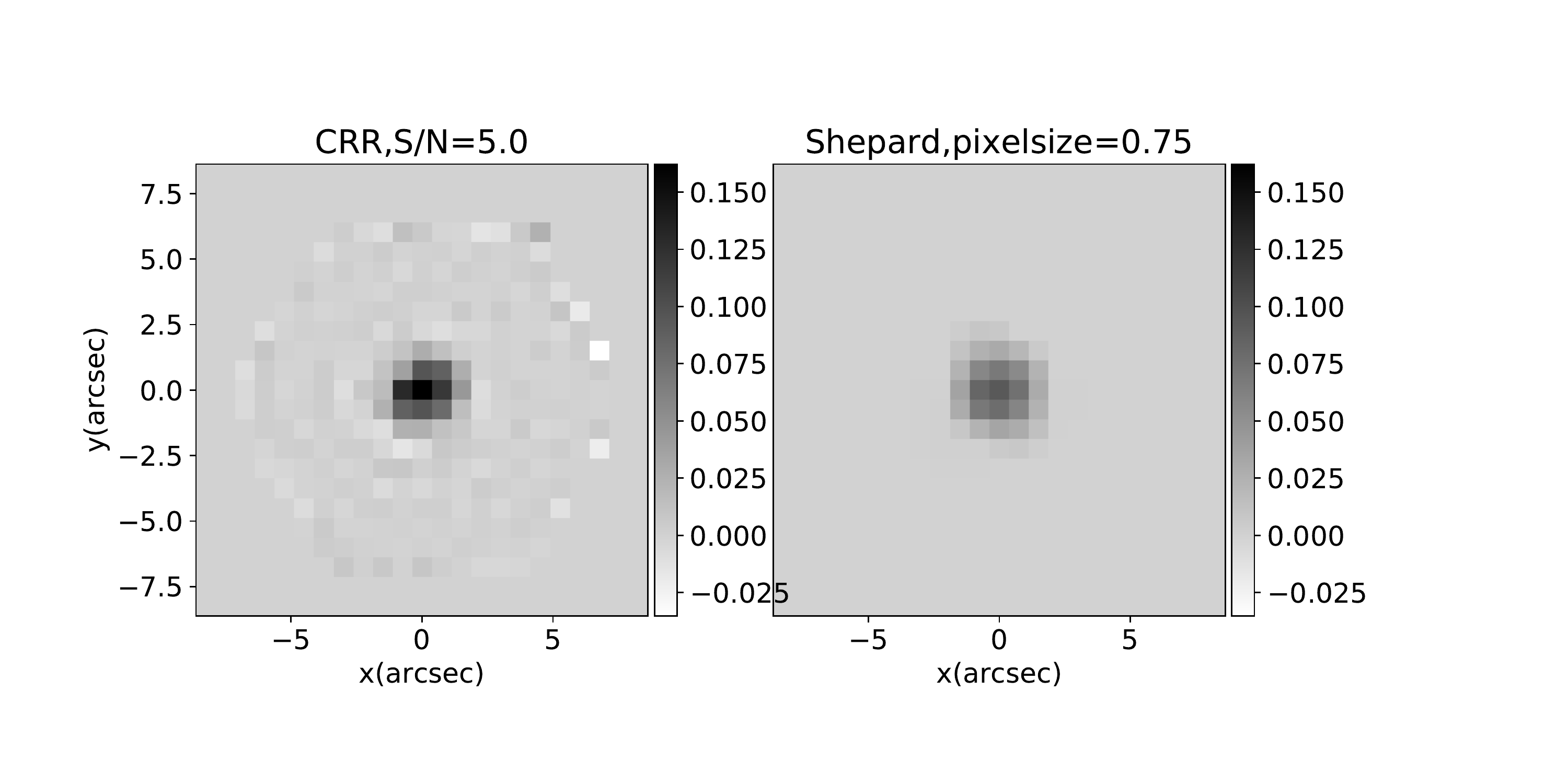}
\caption{Simulated data cube slice with noise at 0.5 
arcsec/pixel and S/N=7.75
for plate-IFU=8720-1901, at  wavelength $\lambda=5500\angstrom$.
Left panel: CRR
method. 
Right panel: Shepard's method.}
\label{dnoise_8720-1901_1}
\end{center}
\end{figure}

Figure  \ref{dnoise_8720-1901_1} shows a point source with
noise, quantified by the signal-to-noise ratio of peak flux in the reconstruction image, for both the 
CRR and Shepard's image.
Obviously the image exhibits some noise. For Shepard's image,
the result looks smoother, due to the larger off-diagonal 
covariances --- i.e. the pixel-to-pixel fluctuations are reduced
because neighboring pixels are correlated.

\begin{figure}
\begin{center}
\includegraphics[trim=2cm 2cm 1cm 0.1cm, width=0.5\textwidth]{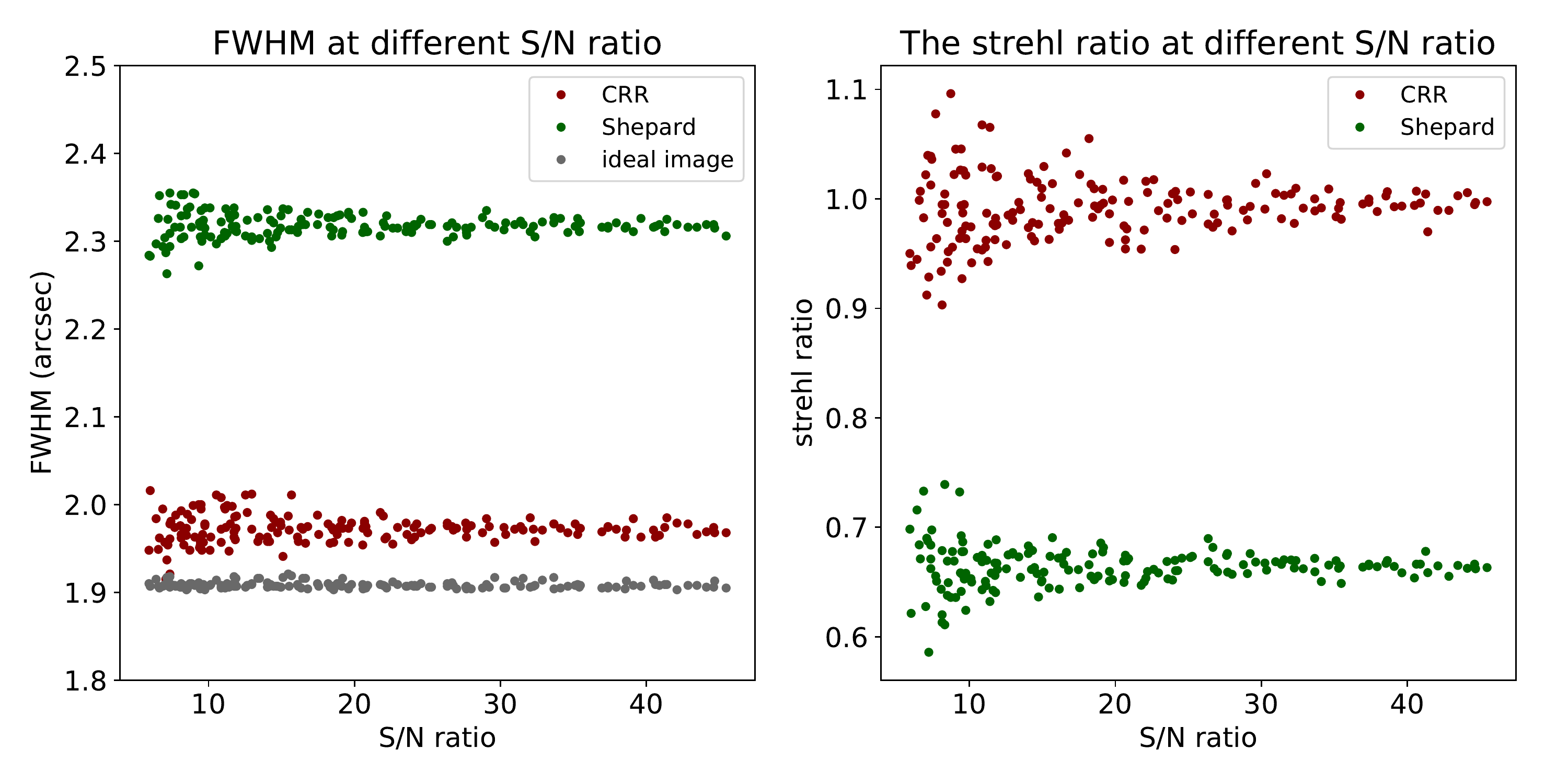}
\caption{Comparison of Shepard's method and CRR method 
at different S/N ratios for plate-IFU=8720-1901 and wavelength
$\lambda=5500\angstrom$. The pixel size=0.75 arcsec. Left panel: FWHM of the PSF. 
Right panel: the pseudo-Strehl ratio of the PSF.}
\label{dnoise_8720-1901_2}
\end{center}
\end{figure}

Figure \ref{dnoise_8720-1901_2} examines the FWHM and 
pseudo-Strehl ratios under a range of noise conditions.
The CRR image remains sharper and brighter in the center
regardless of the noise level. 

\subsection{Effect of an inaccurate PSF model}

We build our analysis based on the assumption that the 
kernel is the same as our assumed one --- that is that 
our double Gaussian model for the seeing with a given
width, convolved with the fiber profile, is correct. 
However, the seeing width is an estimate and its FWHM 
can deviate from reality as much as 20\%; other aspects
of the kernel model may be incorrect as well. 

Shepard's method does not use any information about
the kernel, and thus its behavior is independent of how 
accurately we know the kernel. However, our CRR
method depends on a kernel model to determine the weights. 
Therefore we need to test whether the CRR image varies 
under reasonable assumptions about the inaccuracy of our 
kernel model.

\begin{figure}
\begin{center}
\includegraphics[width=0.5\textwidth]{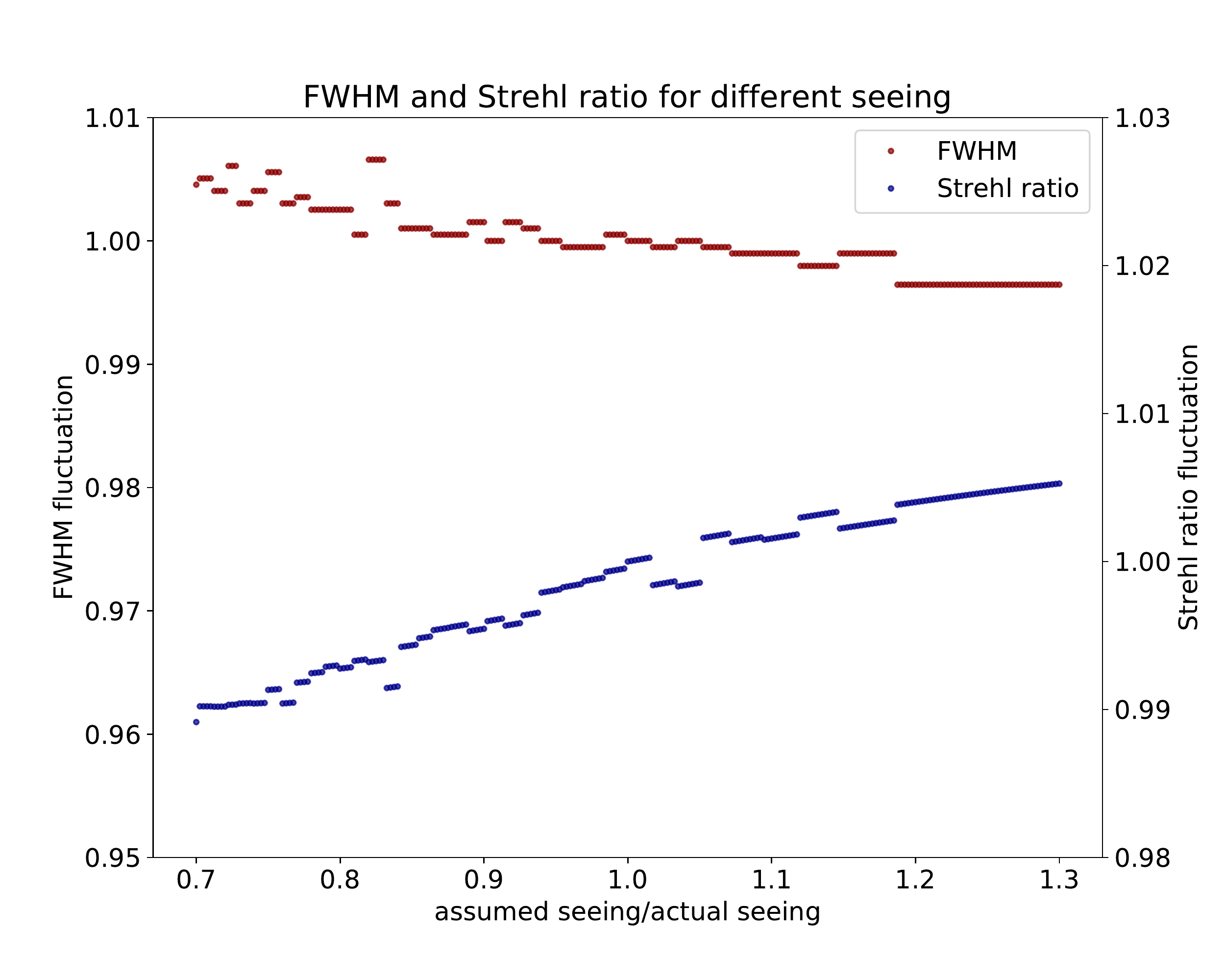}
\caption{The FWHM and Strehl ratio of the CRR image when
the assumed kernel deviates from actual (simulated) kernel, 
for plate-IFU=8720-1901 at wavelength $\lambda=5500\angstrom$, 
with pixel size 0.75 arcsec. The $x$-axis indicates the ratio of 
the assumed FWHM of kernel to the average of the actual seeing 
values for the observations (1.19 arcsec). The piecewise 
discontinuities are an artificial result of the fact that our kernel 
is constructed at a set of discrete FWHM values. }
\label{dPSF_8720-1901_nradial}
\end{center}
\end{figure}

To test the behavior of the algorithm under these conditions, 
we varied the FWHM of the assumed seeing from the actual 
value in our simulations, and compared the CRR image 
with the case that the assumed seeing was the same as 
the actual seeing.  The results are shown in Figure 
\ref{dPSF_8720-1901_nradial}.
We varied the seeing by as much as 30\% from the 
observational estimate for this plate of 1.19 arcsec. The 
FWHM  and pseudo-Strehl ratio for this case are 
1.973 arcsec and 0.996, when the seeing is correctly 
estimated. For a 10\% error in our assumed seeing, 
the FWHM of CRR image changes by around 
0.30\%; for a 20\% error in our assumed seeing, the FWHM of 
CRR image changes by around 0.40\%. The fractional
variation of the Strehl ratio is also similarly small. 
Therefore our results are relatively insensitive to 
whether or not we know the kernel exactly. We have checked
that this variation is small independent of the actual
seeing in the simulation (e.g. if the actual seeing is
poor rather than the relatively good value of 1.19 arcsec).

\subsection{Regularization parameter}

\begin{figure*}
\begin{center}
\includegraphics[width=1\textwidth]{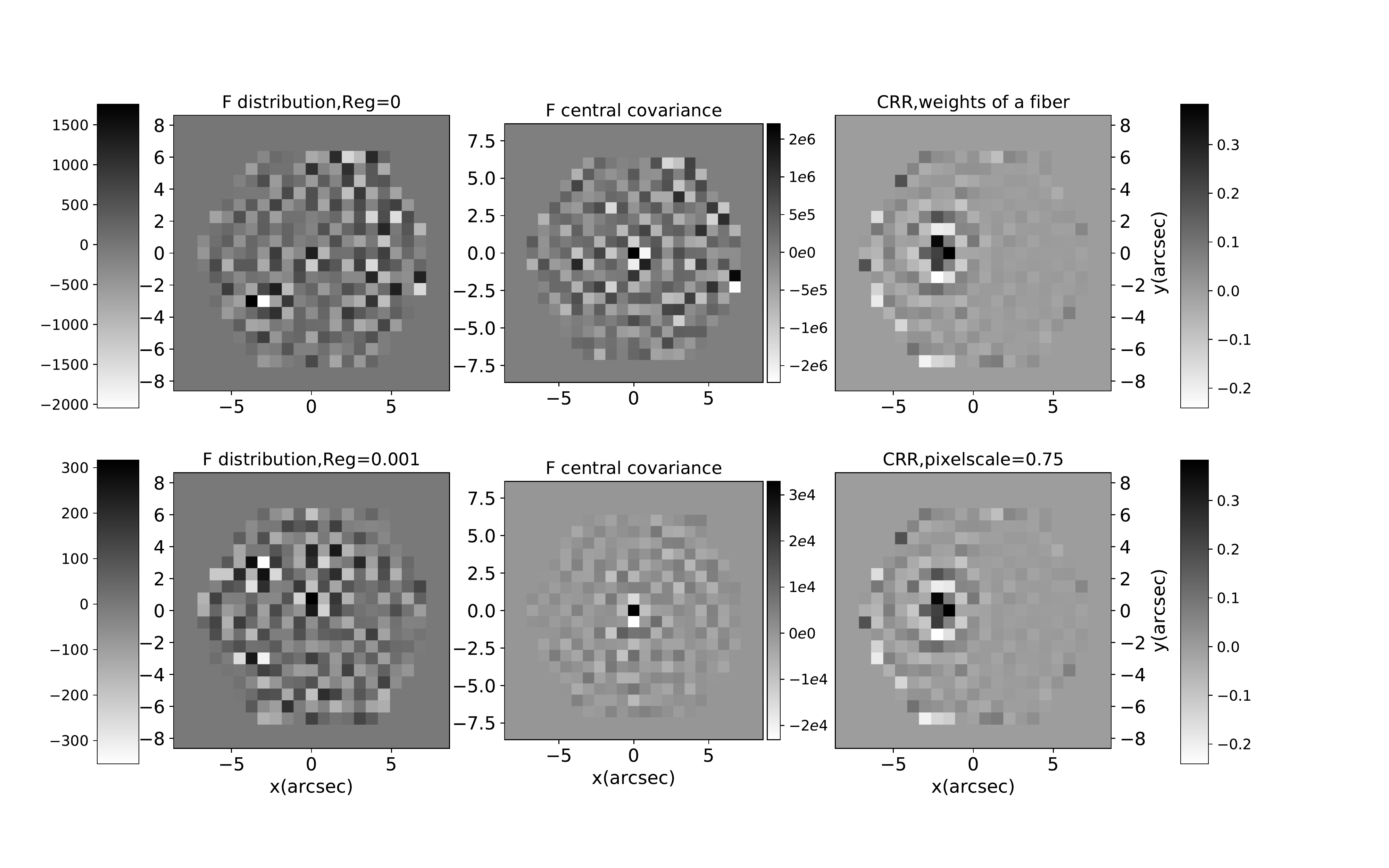}
\vspace*{-10mm}
\caption{The effect of regularization for our method for plate-IFU 
8720-1901. Upper panels are without regularization and lower panels 
are with regularization $10^{-3}$. We used pixel size
0.75 arcsec/pixel. Left panels: the distribution of the 
deconvolved solution. Middle panels: the covariance between the 
central pixel and other pixels in the deconvolved solution. 
Right panels: the contribution of a particular fiber to all pixels 
in the CRR image.}
\label{de_8720-1901_reg}
\end{center}
\end{figure*}

One of the free parameters in the method is the regularization 
parameter $\lambda$. Here we examine the effect of this parameter 
on the results. Figure \ref{de_8720-1901_reg} shows results 
without regularization in the upper panels 
and with regularization in the lower panels, for comparison. 
The left panels show the 
deconvolved reconstruction $\vec{F}$ in the upper left panel 
(using 0.75 arcsec/pixel) for a simulated point source,
showing significant  ringing, with large correlations and 
anti-correlations between pixels. The upper center panel 
quantifies the covariance between the central pixel of $\vec{F}$
and all the other pixels (the central row of the covariance 
matrix), showing large anticorrelations between the central 
pixel and its immediate neighbors. 

These correlations and anticorrelations are large and can introduce
numerical instability to our analysis, since $\vec{F}$ results 
from the last four factors in 
Equation \ref{eq:reconstruction_w}.
However, with appropriate regularization, the behavior of $F$ (and
therefore those last three factors) is much more constrained. While
improving the numerical stability, this regularization will not affect
the extraction of our final reconstruction from the fiber data, due to
the first factor $\mathbf{R}$ in Equation \ref{eq:reconstruction_w}.

The right panels show the contribution of a particular 
fiber to all the pixels on the grid, equivalent to a column 
of the $\mathbf{W}$ matrix. Unlike the Shepard's method weights,
the contributions are not all positive and exhibit some ringing.
This behavior is characteristic of accurate interpolation kernels,
such as sinc-interpolation, so it is expected. These weights
remain identical even under regularization, meaning
that for the values of $\lambda$ we use (or smaller), it 
does not affect the final results.

\subsection{Consistency across wavelength}

The locations of fibers and the kernel shape are a 
function of wavelength, which causes a necessary variation with 
wavelength of the weights $\mathbf{W}$ in our method. Here we examine 
whether the CRR method produces spectra that are consistent 
across wavelength under this variation. 

\begin{figure}
\begin{center}
\includegraphics[width=0.5\textwidth]{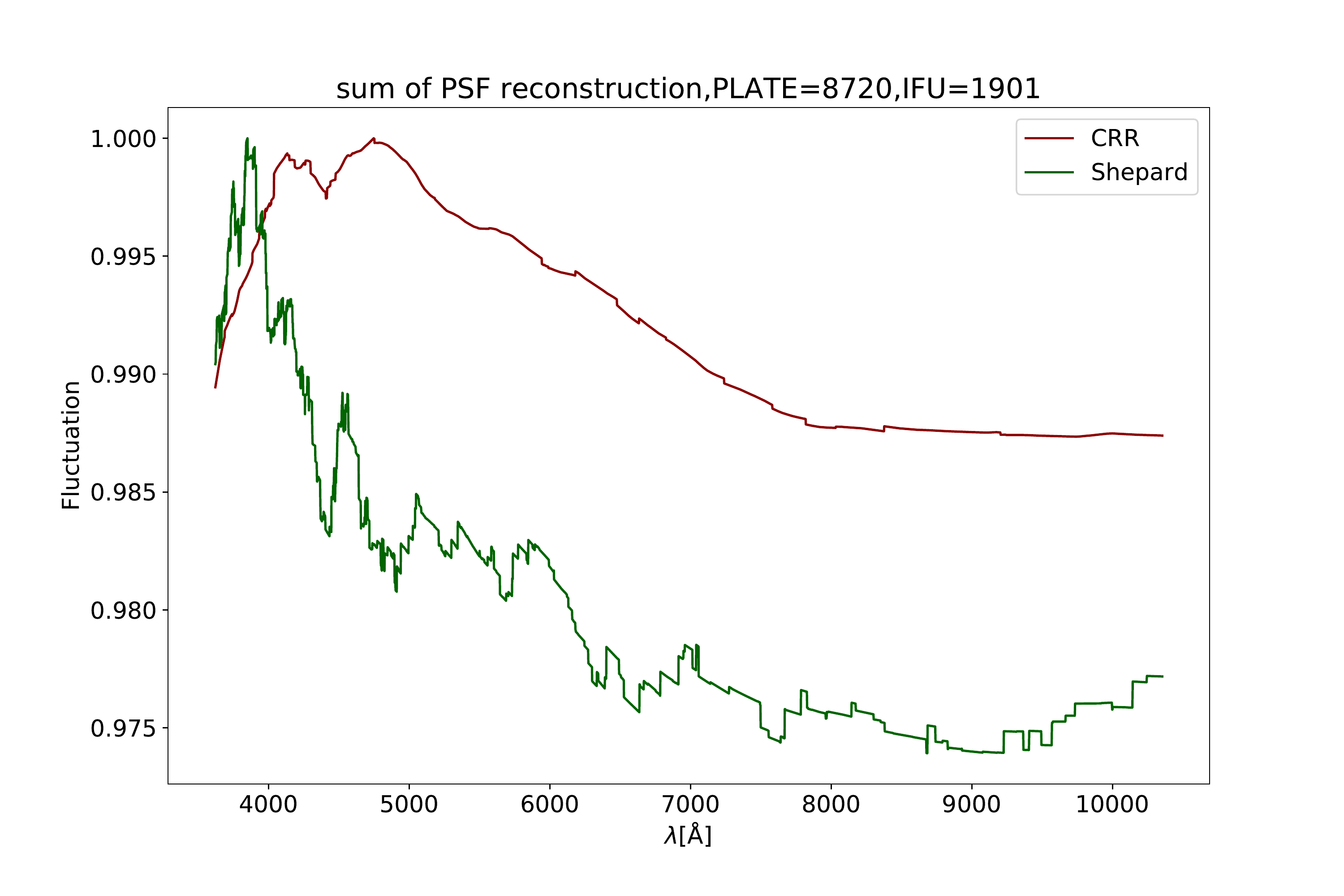}
\caption{Spectrum of the sum of all pixels with for a constant $f_\lambda$ point source for Shepard's method and CRR method. The $y$-axis is normalized by its maximum value.}
\label{dwave_8720-1901_PSFsum}
\end{center}
\end{figure}

\begin{figure}
\begin{center}
\includegraphics[width=0.5\textwidth]{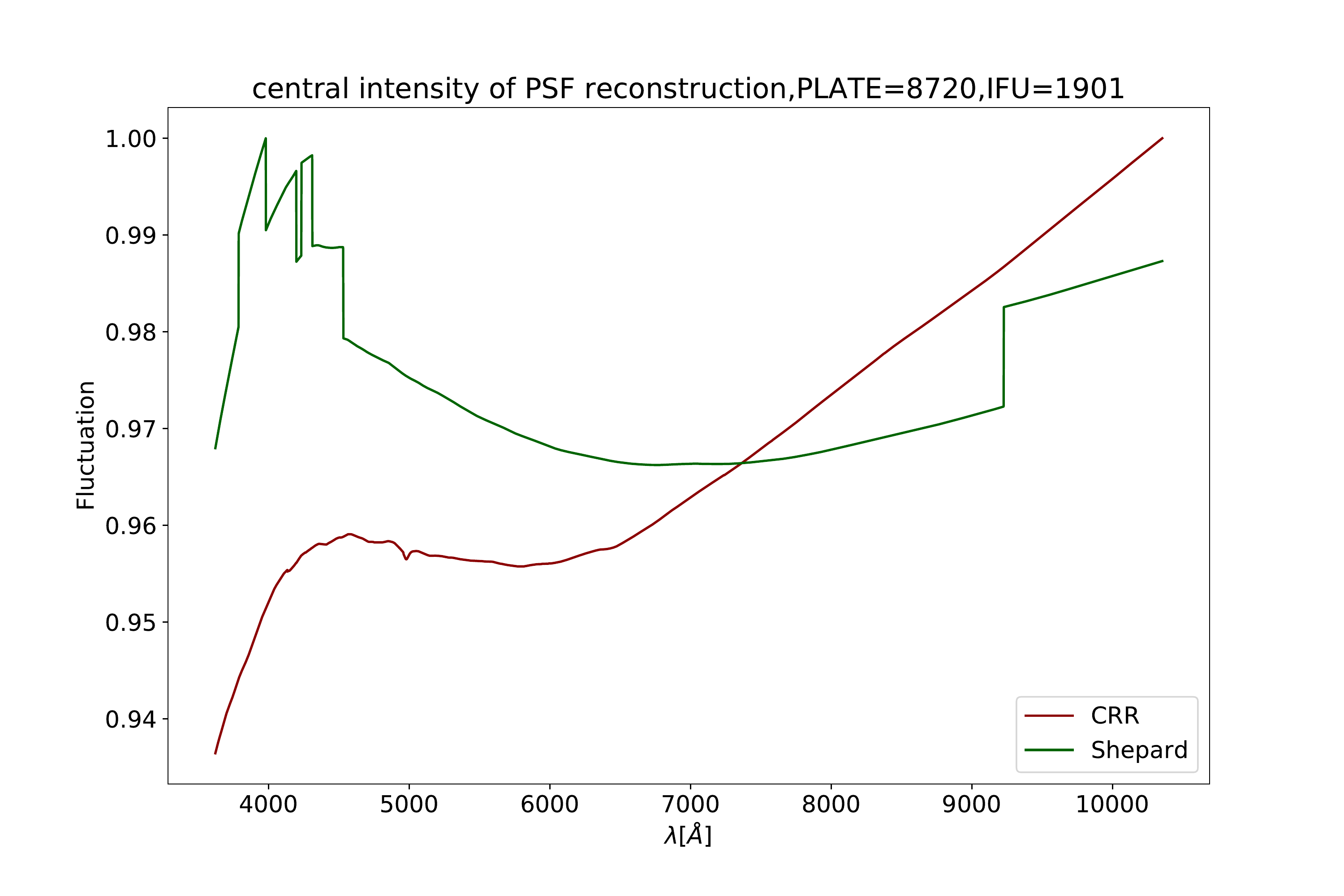}
\caption{Similar to Figure \ref{dwave_8720-1901_PSFsum}, for the spectrum of the central pixel.}
\label{dwave_8720-1901_PSFcenter}
\end{center}
\end{figure}

We simulate a constant $f_\lambda$ point source for all wavelengths 
and use the meta-data (positions, atmosphere conditions) for 
plate-IFU 8720-1901. We compare the output spectra for CRR and 
Shepard's method in Figure \ref{dwave_8720-1901_PSFsum} 
for the total flux within the bundle.
In Figure \ref{dwave_8720-1901_PSFcenter}, we perform the 
same comparison for just the spectrum in the central pixel. 

Figure \ref{dwave_8720-1901_PSFsum} shows that the variation of the sum of 
all the pixels in the simulation results is within about 3\% of constant. 
Once the overall difference in amplitude of $\sim 1\%$ between the
CRR and Shepard's method are accounted for, the relative
variation is consistent between the methods to about 0.5\%. The 
variation in both methods is due to not having enough dithers of 
exposures, which inevitably leads to a variation in 
the flux due to the changing position of the point source 
relative to the fibers as a function of wavelength. Artificially 
adding in more dithers can reduce this variation for the 
simulations (though of course we cannot do that for the observations!).

Figure \ref{dwave_8720-1901_PSFcenter} shows 
the variation of the central pixel intensity across wavelength. The 
variation of intensity is 6\% for the CRR result, compared 
to 3\% for Shepard's result. Therefore, for an individual pixel
there is a slightly larger spectrophotometric inconsistency 
in the CRR method than in Shepard's method. 

For the real case, besides the non-constant flux and extended 
fiber configuration, we need to consider the bad fibers 
identified by the MaNGA DRP for each plate-IFU, leading to
the ``low coverage'' or ``no coverage'' masks. These
fiber masks occur because  of cosmic ray events, bad flat 
fields, CCD defects, broken fibers,  or 2D extraction 
problems \citep{Law2016}. They will result in  zero weights in 
the $\mathbf{W}$ matrix. But we have verified that they do not 
affect the spectrophotometric consistency as a function of 
wavelength, because of the normalization over the contributions 
from all the good fibers.


\begin{figure}
\begin{center}
\includegraphics[width=0.5\textwidth]{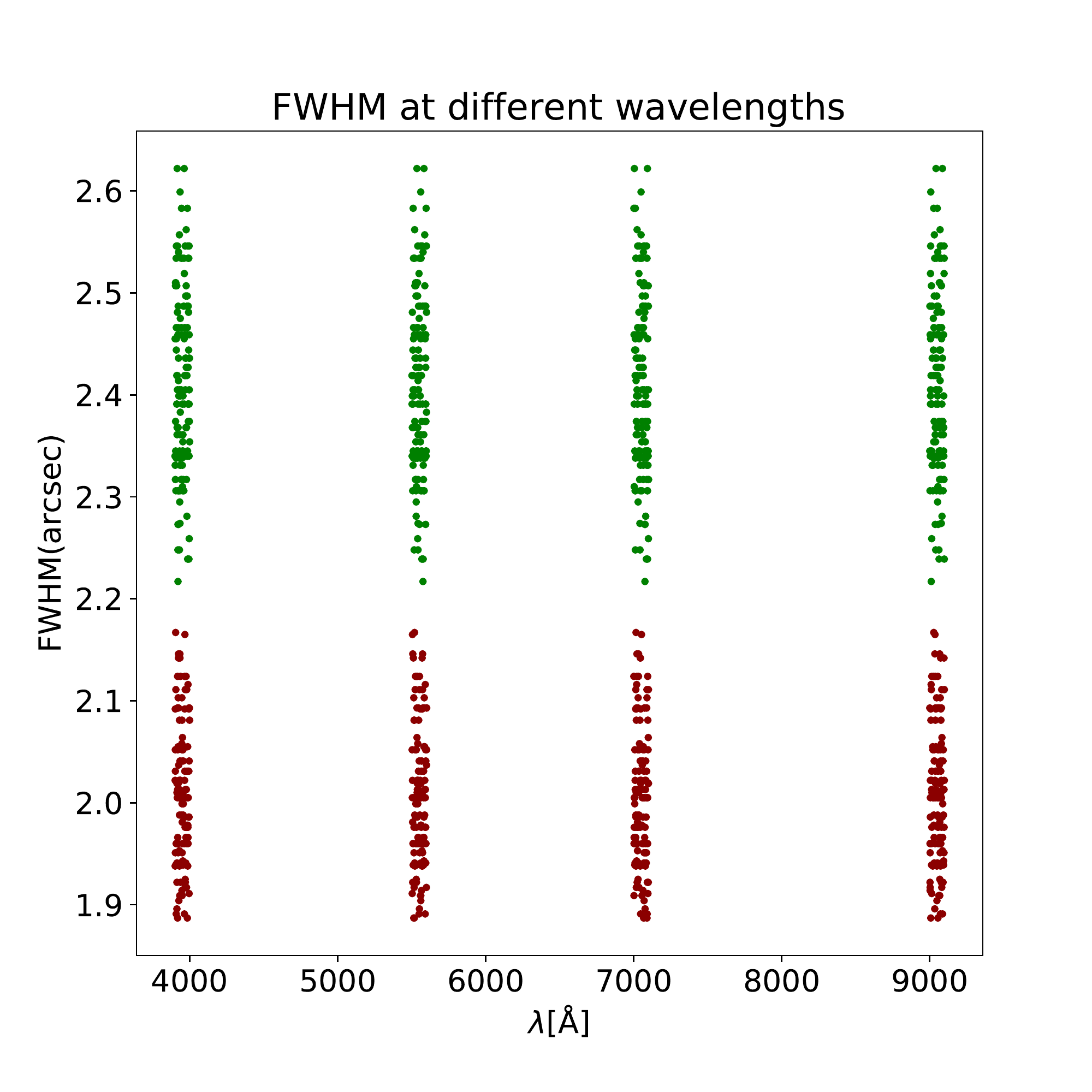}
\caption{PSF FWHM of point source simulation results at several
different wavelengths. Dark red points are for CRR image, and green points are for Shepard's image. 
The points are randomly offset slightly in wavelength for 
clarity. The four wavelength slices we use are
$\lambda=[3900,5500,7000,9000]$ $\angstrom$.}
\label{dreal_FWHM}
\end{center}
\end{figure}

\begin{figure}
\begin{center}
\includegraphics[width=0.5\textwidth]{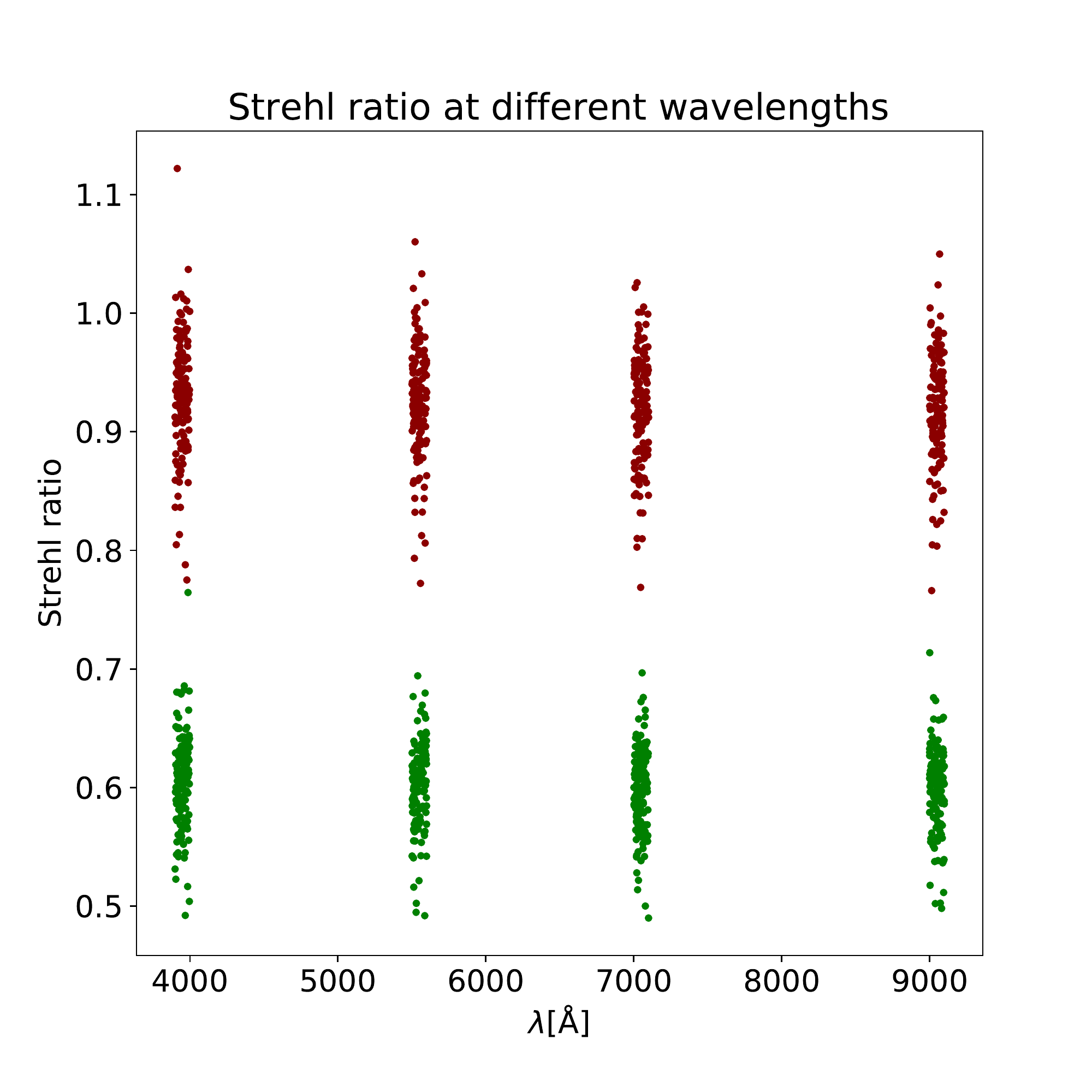}
\caption{Similar to Figure \ref{dreal_strehl}, for the 
pseudo-Strehl ratio of the point source simulation results.}
\label{dreal_strehl}
\end{center}
\end{figure}

We applied our method on 140 fiber bundles in order to demonstrate 
that the example plate-IFU  (8720-1901) is typical. In each case, 
we simulate a point source at the center of the cube. We compare 
the PSF FWHM and central intensity between CRR and 
Shepard's image at several different wavelengths. 
As shown in Figure \ref{dreal_FWHM} and Figure \ref{dreal_strehl}, 
the reconstruction in our method is generally narrower than 
Shepard's result and with higher pseudo-Strehl ratio. 

\section{Demonstration with actual data}
\label{sec:realdata}

The tests above regard a simulated point source.
In this section, we apply the method to several plate-IFUs
in MaNGA as a demonstration and to verify that it is working 
as expected. We will defer a thorough battery of tests of
real data to the next phase of this work.

\begin{figure*}
\begin{center}
\includegraphics[width=1\textwidth]{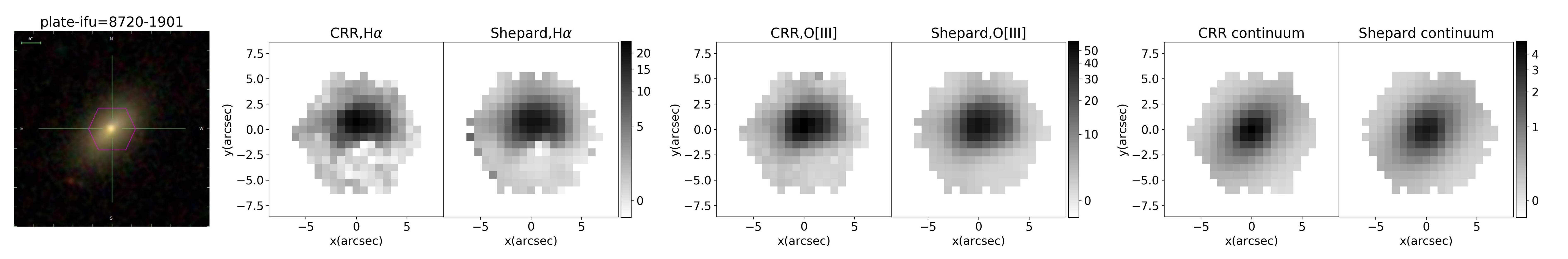}
\includegraphics[width=1\textwidth]{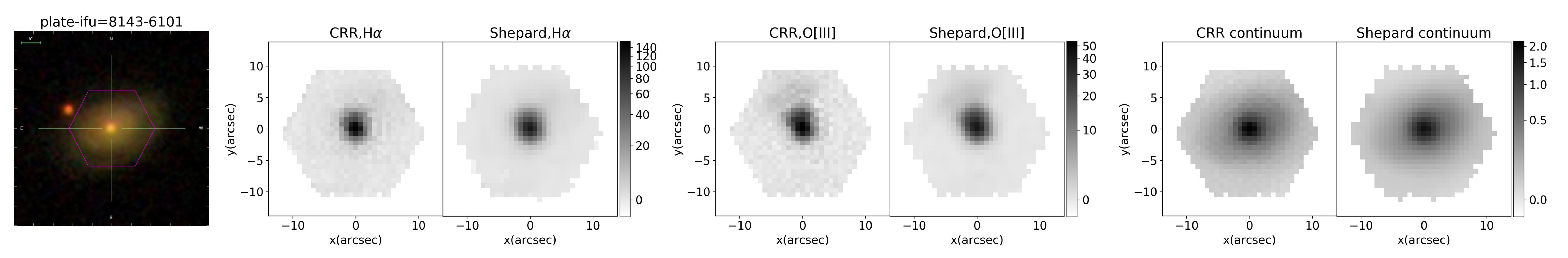}
\includegraphics[width=1\textwidth]{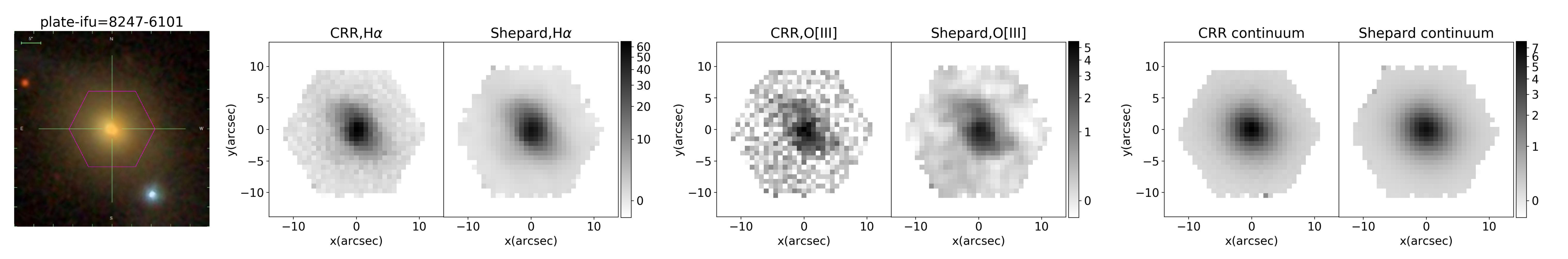}
\includegraphics[width=1\textwidth]{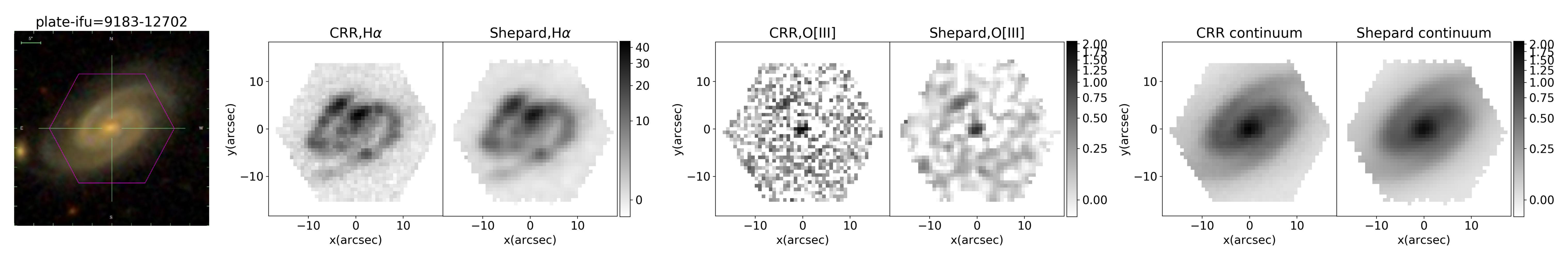}
\caption{Real galaxy extraction slices for four different plate-IFUs.
  From top to bottom, the rows correspond to plate-IFUs 8720-1901,
  8143-6101, 8247-6101 and 9183-12702. From left to right the columns 
  are the $irg$-band color image in
  MaNGA, the continuum subtracted CRR images in H$\alpha$
  6563$\angstrom$ region and O[III] 5007$\angstrom$ region and a
  continuum 5300-6000$\angstrom$ region. In each case, we show our
  method on the left and Shepard's method on the right.}
\label{dreal_HO}
\end{center}
\end{figure*}

Figure \ref{dreal_HO} shows four test galaxies.
As the $irg$-band color images in the left column show, two are
elliptical galaxies and two are spiral galaxies. The first two rows
(plate-IFUs 8720-1901 and 8143-6101) are AGN, and the second two
(plate-IFUs 8247-6101 and 9183-12702) are non-AGN. We use these 
different type of galaxies just to show the variety of possible
cases.

The remaining columns, from left to right, show the H$\alpha$ emission
in the CRR method and in Shepard's method, the [O~III] 5008
emission in both methods, and the mean continuum of the 5300--6000 
\AA\ region in both methods. The H$\alpha$ and [O~III] line emission
fluxes are
estimated by subtracting a continuum estimate using side bands around
each line, and an unweighted integration of the line flux in a fixed
rest frame wavelength range around each line. This relatively crude
method is adequate to characterize the image quality, and we defer to
the next phase of this work a more careful analysis with the MaNGA
Data Analysis Pipeline (\citealt{westfall19a}).

In each image, the sharper nature of CRR image is clear. This
clarity is most dramatically shown in the H$\alpha$ emission in the
bottom row and in the [O~III] emission in the second row. It is also
clear that the off-diagonal covariance in the CRR images is
lower, manifesting as a noisier-looking image. For example, in the
[O~III] images in the bottom two rows, the regions without significant
emission show uniform white noise in the CRR image, but show the
characteristic mottling of correlated noise in Shepard's method.

\begin{figure*}
\begin{center}
\includegraphics[width=1\textwidth]{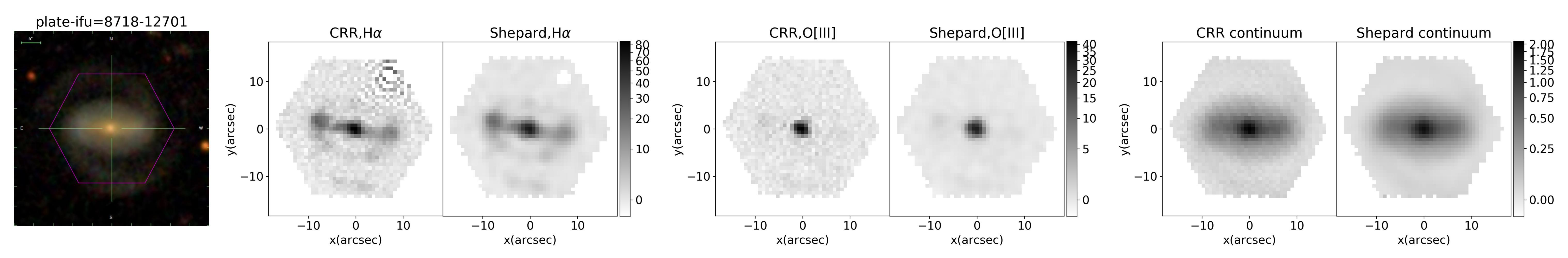}
\includegraphics[width=1\textwidth]{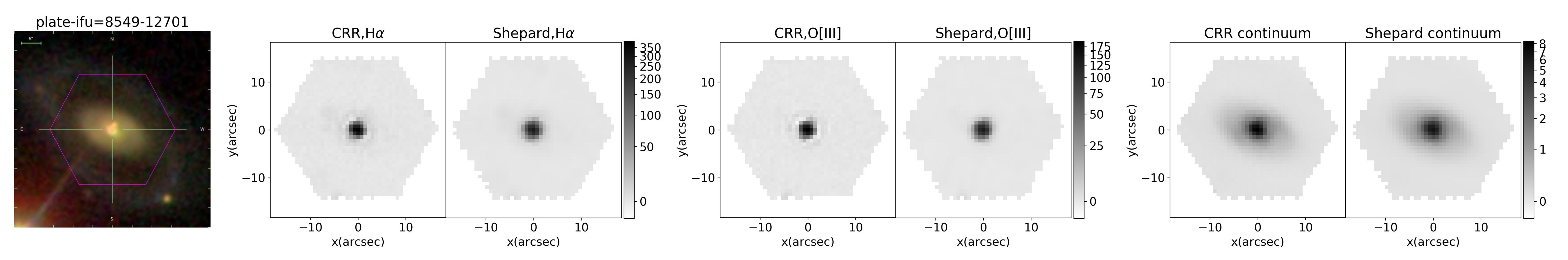}
\includegraphics[width=1\textwidth]{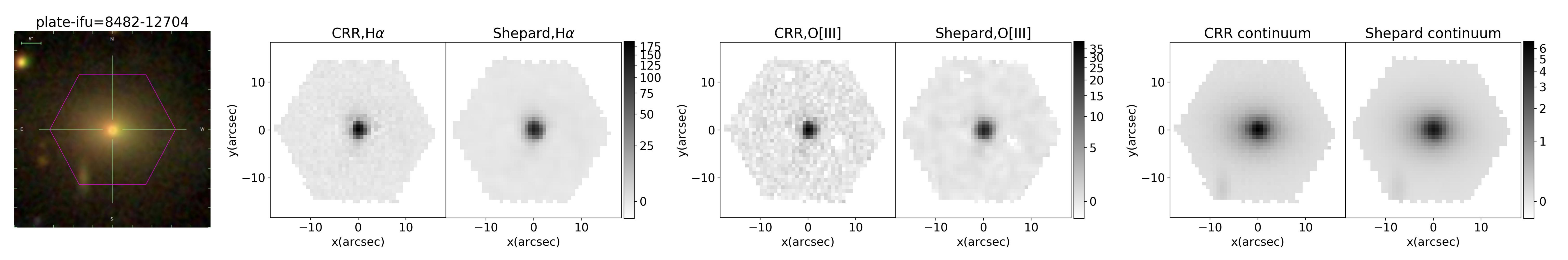}
\caption{Similar to Figure \ref{dreal_HO}, for three AGN-like plate-IFUs: 8718-12701, 8549-12701 and 8482-12704.}
\label{dreal_HO_2}
\end{center}
\end{figure*}

We also check the profile as a function of radius from the center for
real images. Considering that the morphology of the galaxy will affect the radial profile, it is best to choose point-like sources for the comparison. In Figure \ref{dreal_HO_2}, we select three targets identified
as AGN by \citet{Rembold2017}. Using the same methods we use for 
the PSF fitting, we
measure the FWHM and central flux. Table \ref{table_real} shows the
results. Only the comparison of FWHM and central intensity for
different methods has meaning, not the absolute value of these
quantities, since the galaxies are more extended than a PSF and may
have more irregular morphologies than a PSF. In addition, we expect
for extended sources the change in their size to be smaller than for a
PSF. The CRR image reveals a 28.3\% brighter center and a 16.6\% smaller
spatial size on average based on these three cases, showing that the method is working as we expect from the simulation tests in Section \ref{test_nominal}.

\begin{center}
\begin{table}
\caption {Radial profile measurements for real galaxies. FWHM is in
  units of arcsec, while the central flux is in units of
  $10^{-17}$erg/s/cm$^2/\angstrom$/arcsec$^2$.}
\label{table_real}
\begin{tabular}{c|c|c|c|c|c}
\hline
\multirow{2}{*}{\backslashbox{plate}{line}}&\multirow{2}{*}{method}
&\multicolumn{2}{c|}{H$\alpha$ 6563$\angstrom$}&\multicolumn{2}{c}{O[III] 5007$\angstrom$}\\
\cline{3-4} \cline{5-6}&&FWHM&center&FWHM&center\\
\hline
\multirow{2}{*}{8718-12701}&CRR& 2.511&79.5&2.181&42.9\\
&Shepard&2.729&62.7&2.644&29.0\\
\hline
\multirow{2}{*}{8549-12701}&CRR&2.193&405&2.184&201.6\\
&Shepard&2.640&278&2.591&202\\
\hline
\multirow{2}{*}{8482-12704}&CRR&2.283&181&2.264&37.8\\
&Shepard&2.729&128&2.729&26.2\\
\hline
\end{tabular}
\end{table}
\end{center}

\begin{figure}
\begin{center}
\includegraphics[width=0.5\textwidth]{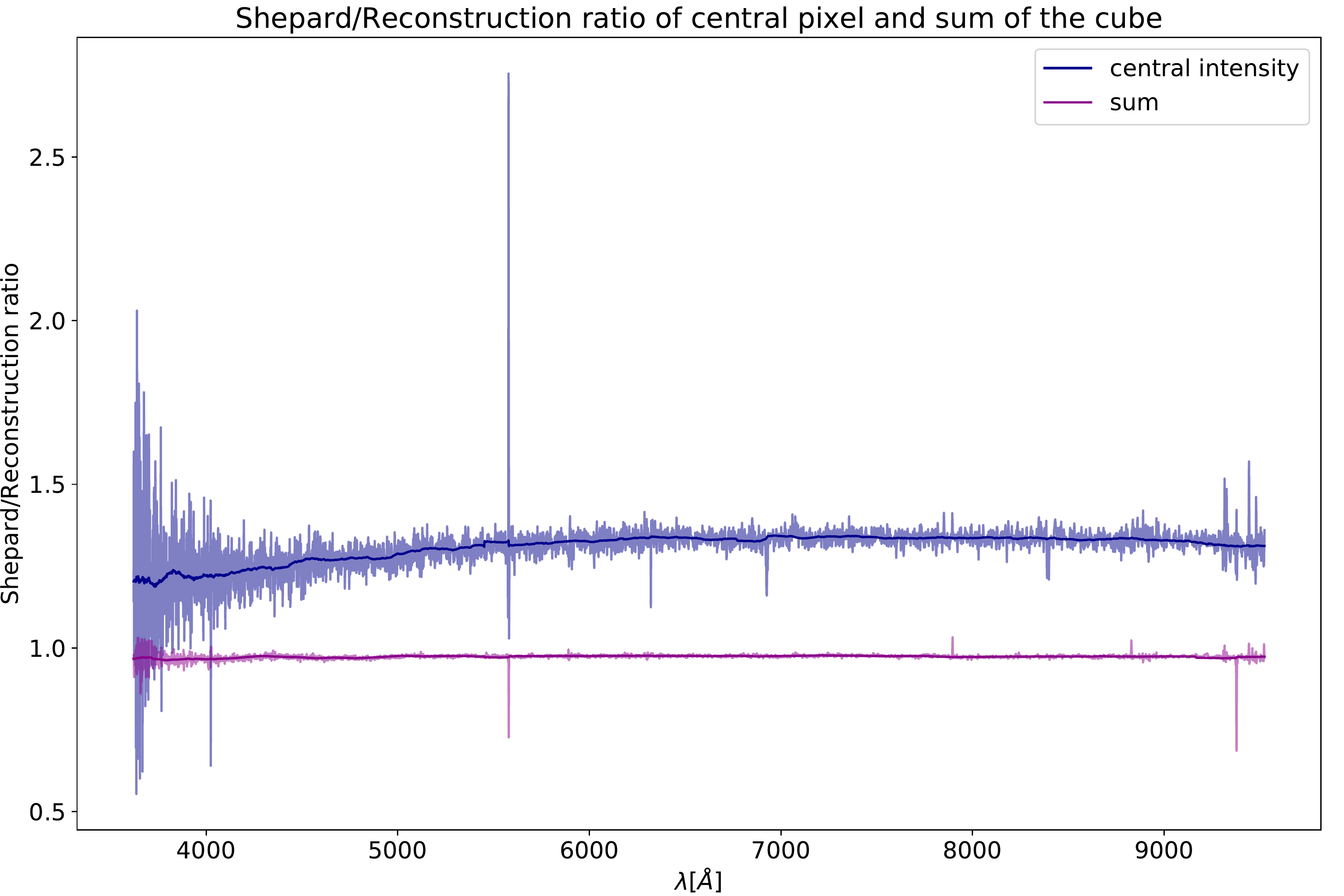}
\caption{Ratio of the spectrum from CRR method to the 
spectrum from Shepard's method for plate-IFU 8720-1901. The pixel scale=0.75 arcsec. 
The blue line is for the central pixel, and the magenta line is 
for the sum of all pixels. 
The solid curves are running average for the ratio smoothing over 100 wavelength slices.}
\label{dwave_8720-1901_cube_ratio}
\end{center}
\end{figure}

Finally, we check the full spectra of the CRR and Shepard's
image results for real data.  Figure \ref{dwave_8720-1901_cube_ratio}
shows the ratio of the spectrum from CRR to the
spectrum from Shepard's image. We consider the full spectrum summed
over all pixels, and the intensity of the central pixel, as labeled,
for plate-IFU 8720-1901. The thick solid curve overlaid on each
spectral ratio is the running mean for the ratio smoothing over 100 wavelength slices or 84$\angstrom$. For the sum of all pixels of the running mean, the ratio of Shepard's image and
our reconstruction is 0.9729 on average with 0.0033 for the standard
deviation, meaning it is almost constant in the running mean. The central
intensity of our reconstruction is 29.9\% brighter than Shepard's result, with
a standard deviation of 0.0439. These results indicate that we have not introduced
major wavelength dependent artifacts relative to what might exist in
Shepard's method.

\section{Discussion}
\label{sec:discussion}

Shepard's method is one of the most commonly used 
techniques to  interpolate irregularly sampled data 
onto a regular grid (see \citealt{franke80a, dellaccio16a}). 
The largest IFU surveys to date are all using a variant
of it. MaNGA and CALIFA both used the form of Shepard's
method described here, and SAMI uses a differently motivated 
technique, but one which is very similar to 
Shepard's method with a different kernel.

However, Shepard's method produces images whose flux 
errors are correlated with one another; i.e. they 
have a very non-diagonal covariance matrix (\citealt{Law2016}). 
In addition, as we found in this investigation, it 
unnecessarily broadens the PSF of the resulting image.

Other techniques exist in the statistical literature 
for interpolating from irregularly distribution samples,
for example radial basis function techniques,
Wiener interpolation, ``kriging,'' and Gaussian
processes (\citealt{krige51a, wiener64a, schaback95a,
hartkamp99a, press07a, rasmussen06a}). However, these methods
are not designed to produce a consistent and tight PSF,
and generally lead to highly off-diagonal covariances.

Motivated by a desire to avoid off-diagonal covariances,
we examined the techniques of \citet{Bolton2010}, and
in this paper have adapted them to the imaging context. 
The result is a technique which successfully reduces the
off-diagonal covariances to a very small level and also
provides a final image PSF which is better than other
methods by around 16\% in the FWHM, for the MaNGA 
example we consider. 

The off-diagonal covariances produced by other methods
can heavily affect the subsequent analysis. For 
simple measurements such as aperture fluxes, the 
correct propagation of errors becomes cumbersome and complex. 
For more complicated measurements such as maximum 
likelihood model fitting, even determining the best
fit parameters depends on accounting for the covariance
accurately. The MaNGA data, and IFU data generally, is often
used for such measurements. Our reconstruction
represents a way to simplify these measurements up front.

One may wonder why it is possible for us to obtain near-zero
off-diagonal covariances, considering that dithered samples 
are spaced every 1.44'' while our pixel scale is 0.75''. 
First, it is worth noting that the covariance between adjacent
neighbors is not entirely eliminated, but remains at the few percent level 
for this pixel size. Second, the differences in seeing and even
slight differences in coordinates for different exposures 
allow the method to create nearly independent data points 
in neighboring pixels.

The improved image PSF is substantial. It is equivalent to building
an instrument with 30--40\% more fibers. It provides a
greater ability to resolve structures within galaxies and to measure
gradients accurately, as well as to find fainter point-like features
(e.g. AGN).

A critical aspect of obtaining diagonal covariances is that some of
the weights are negative, whereas Shepard's method has all positive
weights (see Figure \ref{de_8720-1901_contribution}). All positive weights will always produce correlated
errors. Although at first glance the fact that a positive fiber flux
can contribute negatively to a pixel flux may be non-intuitive, this
same feature exists in all accurate image interpolation techniques on
a grid (e.g. those based on a sinc kernel) and is not in itself a
cause for concern.

We plan to apply this method to the entire MaNGA sample 
and to test the MaNGA Data Analysis Pipeline code on 
the revised cubes. Since our method is still experimental 
(we have only applied it to around 100 or so cubes) we 
cannot determine yet if it will yield a practical improvement
to the results, but we view its prospects as promising.

So far as we can determine, although our technique was 
derived from that of \citet{Bolton2010}, this application 
is entirely new, and represents a new method for scattered image 
interpolation. Since its main function is to control the 
covariance matrix of the result, rather than to regularize 
the result by enforcing a notion of smoothness, we refer to
it as covariance-regularized reconstruction.
It is relevant when the input samples are
noisy and irregular, when result of the interpolation is 
meant to be  a specific grid of values, and when there is 
a natural  resolution (in our case, the kernel) in the 
sampled  image that is meant to be preserved. These 
conditions apply in other IFU data sets, as well as to 
a number of ground-based and space-based imaging data sets.
Thus, this method may provide an alternative and improved
method to analyze those data sets.

\section*{Acknowledgements}

We thank Yacine Ali-Ha{\"i}moud, Stephen Bailey, Adam Bolton, Cristina
Mondino,
and 
David W.~Hogg for useful discussions regarding this work.
MRB was supported in part by National Science Foundation 
grant NSF-AST-1615997.

Funding for the Sloan Digital Sky Survey IV has been provided 
by the Alfred P. Sloan Foundation, the U.S. Department of 
Energy Office of Science, and the Participating Institutions. 
SDSS-IV acknowledges
support and resources from the Center for High-Performance 
Computing at
the University of Utah. The SDSS web site is www.sdss.org.

SDSS-IV is managed by the Astrophysical Research Consortium for the 
Participating Institutions of the SDSS Collaboration including the 
Brazilian Participation Group, the Carnegie Institution for Science, 
Carnegie Mellon University, the Chilean Participation Group, the French Participation Group, Harvard-Smithsonian Center for Astrophysics, 
Instituto de Astrof\'isica de Canarias, The Johns Hopkins University, 
Kavli Institute for the Physics and Mathematics of the Universe (IPMU) / 
University of Tokyo, the Korean Participation Group, Lawrence Berkeley National Laboratory, 
Leibniz Institut f\"ur Astrophysik Potsdam (AIP),  
Max-Planck-Institut f\"ur Astronomie (MPIA Heidelberg), 
Max-Planck-Institut f\"ur Astrophysik (MPA Garching), 
Max-Planck-Institut f\"ur Extraterrestrische Physik (MPE), 
National Astronomical Observatories of China, New Mexico 
State University, 
New York University, University of Notre Dame, 
Observat\'ario Nacional / MCTI, The Ohio State University, 
Pennsylvania State University, Shanghai Astronomical Observatory, 
United Kingdom Participation Group,
Universidad Nacional Aut\'onoma de M\'exico, University of Arizona, 
University of Colorado Boulder, University of Oxford, 
University of Portsmouth, 
University of Utah, University of Virginia, 
University of Washington, University of Wisconsin, 
Vanderbilt University, and Yale University.









\bsp	
\label{lastpage}
\end{document}